\documentclass[prb,twocolumn,showpacs,preprintnumbers,amsmath,amssymb]{revtex4}
\usepackage{color}
\usepackage{graphicx}
\usepackage{amsmath}
\usepackage{overpic}
\usepackage{amsbsy}
\usepackage{dcolumn}
\usepackage{bm}

\begin{document}

\title{Bound states at the interface between antiferromagnets and
superconductors}

\author{Brian M. Andersen$^1$, I. V. Bobkova$^2$, P. J. Hirschfeld$^1$, and
Yu. S. Barash$^2$} \affiliation{$^1$Department of Physics,
University of Florida, Gainesville, Florida 32611-8440,
USA\\$^2$Institute of Solid State Physics, Chernogolovka, Moscow
reg., 142432, Russia}

\date{\today}

\begin{abstract}
We present a detailed theoretical investigation of interfaces and
junctions involving itinerant antiferromagnets. By solving the
Bogoliubov-de Gennes equations with a tight-binding model on a
square lattice, we study both the self-consistent order parameter
fields proximate to interfaces between antiferromagnets (AF) and
$s$-wave (sSC) or $d$-wave (dSC) superconductors, the dispersion
of quasiparticle subgap states at interfaces and interlayers, and
the local density of states (LDOS) as a function of distance from
the interface. In addition, we present the quasiclassical approach
to interfaces and junctions involving itinerant antiferromagnets
developed in an earlier paper. Analytical results are in excellent
agreement with what we obtain numerically. Strong effects of pair
breaking in the presence of low-energy interface Andreev states
are found in particular for AF/sSC interfaces when interface
potentials are not too high. Potential barriers induce additional
extrema in the dispersive quasiparticle spectra with corresponding
peaks in the LDOS. Discrete quasiparticle dispersive levels in AF
- normal metal (N) - AF systems are found to strongly depend on
the misorientation angle of the magnetizations in the two
antiferromagnets.
\end{abstract}


\maketitle

\section{Introduction}
Interfaces of magnetic materials with normal metals and
superconductors have attracted much attention in recent years
because they can strongly influence properties of mesoscopic and
nanoscopic systems, and may play important role in compounds with
competing magnetic and superconducting ordering. Hybrid
superconducting systems involving ferro- and/or antiferromagnets
manifest unusual properties associated with spin and orbital
effects, and are of both fundamental interest and important for
technological applications. Ferromagnetic layers can spin polarize
quasiparticle currents and Zeeman split surface densities of
states, with possible applications in spintronics \cite{sarma04}.
Superconductor - ferromagnet -superconductor (SC/F/SC) junctions
have been shown to display $0-\pi$ transitions with varying
temperature, width, or orientational structure of the
magnetization of the ferromagnetic interlayer
\cite{ryazanov01,aprili02,aprili03,aprili04,buz82,buz92,nazarov01,bb02,bb01}.

There are also many situations of fundamental and practical
interest which involve interfaces with antiferromagnets. In
particular, many of the properties of high-temperature
superconducting (HTS) cuprate materials are thought to result from
a competition between antiferromagnetic and superconducting order,
and there are many naturally occurring situations and possible
devices which might involve such boundaries. These include
interfaces of insulating and highly doped cuprates or
superconductor - antiferromagnet - superconductor (SC/AF/SC)
junctions\cite{jc1,jc2}, HTS grain boundaries\cite{grainreview}
where antiferromagnetism may play a role as a surface state, and
the antiferromagnetism which has been observed in HTS vortex cores
\cite{AFcores1,AFcores2,AFcores3,AFcores4,AFcores5}. At the same
time, there exist only very preliminary results of experimental
and theoretical investigations of proximity and Josephson effects
through various types of antiferromagnetic interfaces
\cite{demler98,zabel99,kanoda02,blamire03,gorkov01,andersen02,bobkova05}.
Below we study theoretically interfaces between itinerant
antiferromagnets and normal metals or superconductors. Itinerant
antiferromagnets like chromium and its alloys
\cite{fawcett88,fawcett94,zabel99} are metals above the N\'eel
temperature. In the antiferromagnetic phase, however, an energy
gap in the quasiparticle spectrum arises either on the whole Fermi
surface or on parts of it. Similar properties are also manifested
in Mott antiferromagnets, in particular undoped cuprates. Since
they possess strong correlations due to large on-site Coulomb
repulsion $U$, the mean-field approach of the present paper cannot
be applied quantitatively to Mott systems, whereas it applies well
to itinerant antiferromagnets with comparatively small $U$. We
expect, however, that our main conclusions regarding AF/S and AF/N
interfaces can be qualitatively applied also to interfaces with
Mott antiferromagnets because they are based largely on symmetry
properties and general characteristics of antiferromagnets like
the doubling of period, nesting conditions and the wave vector of
the antiferromagnetic pattern. In itinerant antiferromagnets, the
energy gap in the quasiparticle spectrum is determined by the
antiferromagnetic order parameter, i.e. the sublattice electronic
magnetization $m$. This applies both to commensurate
antiferromagnetic phases and to phases with spin-density waves
\cite{overhauser62}, and is reminiscent of the situation in
superconductors, where the energy gap is determined by the
superconducting order parameter $\Delta$.

Recently, it has been demonstrated theoretically that a new
spin-dependent channel of quasiparticle reflection, the so-called
$Q$ reflection, takes place at interfaces between itinerant
antiferromagnets and normal metals or superconductors
\cite{bobkova05}. Parallel to the interface, the momentum
component of low-energy normal-metal quasiparticles changes by a
wave-vector $\bm Q_{\|}$ in a $Q$ reflection event, where $\bm Q$
is the wave-vector of the antiferromagnetic pattern. Assuming
small Fermi velocity mismatches and taking into account the
nesting condition $E_F(\bm p+\bm Q)=-E_F(\bm p)$ in itinerant
antiferromagnets, one can see that a normal-metal quasiparticle
changes its total momentum by $\bm Q$ and the respective velocity
changes its sign in a $Q$ reflection event. Hence, normal-metal
quasiparticles with energies less than or comparable to the
antiferromagnetic gap experience spin-dependent retroreflection at
antiferromagnet-normal metal (AF/N) transparent interfaces.
Furthermore, $Q$ reflection processes generate quasiparticle bound
states below the AF gap in AF/N/AF junctions, analogously to the
case of subgap states in SC/N/SC systems formed by Andreev
reflection. The AF/N/AF bound states arise from a coherent
superposition of electrons with momenta ${\mathbf{k}}$ and
${\mathbf{k}}+{\mathbf{Q}}$ and almost opposite velocities. Subgap
states arise also at AF/SC interfaces as a combined effect of
Andreev and $Q$ reflections. Among a variety of subgap states,
low-energy states with energies $E_B\ll {\rm min}\{m,\Delta\}$ are
of special interest since they can result in low-temperature
anomalies in the Josephson critical current, as well as low-bias
anomalies in the conductance. Low-energy quasiparticle interface
states were also predicted to occur on antiferromagnetic -
$s$-wave superconductor (AF/sSC) interfaces in the absence of
specular reflection, when one can disregard effects of interface
potential barriers and Fermi velocity mismatches. For an
sSC/AF/sSC junction, these bound states are split due to a finite
width of the AF interlayer and carry the supercurrent. At AF/dSC
interfaces, low-energy bound states $E_B\ll {\rm min}\{m,\Delta\}$
do not exist, at least if the order parameters are small compared
with the hopping matrix element ($\Delta, m\ll t$). This is
contrary to the case of a (110) surface of a dSC confined with
impenetrable potential wall where zero-energy surface Andreev
states are formed \cite{zesref1,zesref2,zesref3}.

Below, we extend the study of effects of $Q$ reflection processes
based on self-consistent solutions of the Bogoliubov-de Gennes
(BdG) equations. This goes beyond the framework of the preceding
paper \cite{bobkova05}. In general, we find excellent agreement
with the results by Bobkova {\sl et al} \cite{bobkova05}. At the
same time, the more general approach of the present paper allows
us to study several important new problems. In particular, we
discuss the effects of interface potentials on the dispersive
quasiparticle interface states and the LDOS. In the presence of
potential barriers and Fermi velocity mismatches, there exists an
interplay of specular and $Q$ reflections. We demonstrate that
potential barriers on interfaces between AF and either $s$-wave or
$d$-wave superconductors can result in new extrema in dispersive
quasiparticle spectra and additional associated peaks in the LDOS.
We also find new interface quasiparticle states with subgap
energies near the edge of the continuum, which arise due to
self-consistent suppression of the order parameters near the
interface. By studying effects of interface pair breaking at AF/SC
interfaces, we find that for the (110) orientation the
self-consistent suppression of both antiferromagnetic and
superconducting order parameters near the interface is accompanied
by even-odd spatial oscillations. We show that discrete
quasiparticle dispersive levels in AF-N-AF systems strongly depend
on the relative orientation of the magnetizations in the two
antiferromagnets. Effects of the misorientation angle turn out to
be analogous to the influence of the phase difference on the
discrete quasiparticle spectrum in SC-N-SC systems.

The paper is organized as follows.  In Section II, we introduce
the microscopic model used to study various interfaces with
antiferromagnets, the BdG equations with a mean-field treatment of
both magnetism and superconductivity general enough to study both
$s$- and $d$-wave pairing symmetry, as well as various interface
potentials. In Section III, we sketch the derivation of the
associated quasiclassical (Andreev) equations, complemented with
boundary conditions, by assuming slow spatial variations of both
order parameters. In Section IV, we study quasiparticle states at
(100) and (110) interfaces of both $s$- and $d$-wave
superconductors with antiferromagnets, and compare the results of
numerical evaluations of the BdG equations with the predictions of
the quasiclassical theory. We end section IV by showing how the
bound state energies can be obtained within a transfer matrix
formalism. In Section V, we study the AF/N/AF junction, and
discuss a novel ``spin-$\pi$" configuration where the relative
phase of the staggered magnetization on both sides of the junction
can tune the energy of the interface bound states. Conclusions and
perspectives for future work are presented in Section VI.

\section{Model}
For studying the electronic structure of interfaces between
antiferromagnets and superconductors or normal metals, we consider
the following 2D mean-field Hamiltonian on a square lattice
\begin{eqnarray}\label{hamiltonian}
\hat{H}= &-& t \sum_{\langle ij \rangle\sigma}
\hat{c}_{i\sigma}^{\dagger}\hat{c}_{j\sigma} + \sum_{\langle ij
\rangle} \left( \Delta_{ij}
\hat{c}_{i\uparrow}^{\dagger}\hat{c}_{j\downarrow}^{\dagger} +
\mbox{H.c.} \right) \nonumber\\ &-& \sum_{i\sigma} (\mu - h_i)
\hat{n}_{i\sigma} + \sum_{i} m_i \left(\hat{n}_{i\uparrow} -
\hat{n}_{i\downarrow} \right).
\end{eqnarray}
Here, $\Delta_{ij}$ and $m_i$ denote the superconducting and
magnetic order parameters, respectively.
$\hat{c}_{i\sigma}^{\dagger}$ creates an electron of spin $\sigma$
on site $i$, $t$ denotes the nearest neighbor hopping integral,
$\mu$ is the filling factor, and
$\hat{n}_{i\sigma}=\hat{c}_{i\sigma}^{\dagger}\hat{c}_{i\sigma}$
is the particle number operator on site $i$. In
Eq.(\ref{hamiltonian}), $h_i$ is an interface potential. We will
study self-consistently only singlet $s$-wave or $d$-wave
superconducting pairings defined as $\Delta_{ij}=-(V_i/2)\langle
\hat{c}_{i\downarrow}\hat{c}_{j\uparrow} -\hat{c}_{i\uparrow}
\hat{c}_{j\downarrow} \rangle$. For $s$-wave pairing, one should
put $i=j$, whereas the $d$-wave order parameter $\Delta_{ij}$
connects nearest neighbor sites. The self-consistent magnetic
order parameter is represented as $m_i=(U_i/2) \left[ \langle
\hat{n}_{i\uparrow} \rangle\ - \langle \hat{n}_{i\downarrow}
\rangle \right]$. In the bulk of the antiferromagnet and in the
absence of any perturbations, the staggered magnetic order
parameter takes the form $m_j=(-1)^{j_a+j_b}m=\exp(i\bm Q \bm
j)m$. For a square lattice with the crystal coordinate axes $a$
and $b$, the antiferromagnetic wave vector is $\bm
Q=(\pi/a,\pi/a)$. Within the framework of a generic Hubbard-like
model, the staggered antiferromagnetic gapped state is stable only
at or near half filling. For this reason we assume below vanishing
or small $\mu$.

We choose a coordinate system where $x$ and $y$ describe
coordinates perpendicular and parallel to the interface,
respectively. For a (100) interface the $x$ and $y$ axes coincide
with the crystal axes $a$ and $b$. Then the normal-state electron
band is given by
\begin{equation}
\xi({\bf k})=- 2 t (\cos k_a +\cos k_b) - \mu,
\label{xi1_100}
\end{equation}
and the respective Brillouin zone is spanned by $k_{a,b} \in
[-\pi,\pi]$, with the momenta given in units of $a^{-1}$. For a
(110) interface we have instead
\begin{equation}
\xi({\bf k})=- 4t\cos\bigl(k_x/\sqrt{2}\bigr)
\cos\bigl(k_y/\sqrt{2}\bigr) - \mu,
\label{xi1_110}
\end{equation}
with $k_{x}\in[-\sqrt{2}\pi,\sqrt{2}\pi]$,\, and
$k_{y}\in[-\pi/\sqrt{2},\pi/\sqrt{2}]$, on account of the periodic
conditions along the crystal surface.

The Hamiltonian (\ref{hamiltonian}) is quadratic in the Fermi
fields and can be diagonalized with Bogoliubov transformations,
${\hat{c}_{i\sigma}}^\dagger = \sum_{n} (u^*_{n\sigma}(i)
\hat{\gamma}_{n\sigma}^\dagger + \sigma v_{n\sigma}(i)
{\hat{\gamma}}_{n\overline{\sigma}})$. The corresponding
Bogoliubov-de Gennes equations take the form
\begin{equation}\label{BdG}
\sum_j \left( \begin{array}{cc} {\mathcal{K}}^{+}_{ij,\sigma}&
{\mathcal{D}}_{ij,\sigma} \\
{\mathcal{D}}^*_{ij,\sigma} & -{\mathcal{K}}^{-}_{ij,\sigma}
\end{array} \right) \left( \begin{array}{c} u_{n\sigma}(j) \\
v_{n\overline{\sigma}}(j) \end{array}\right) = E_{n\sigma} \left(
\begin{array}{c} u_{n\sigma}(i) \\
v_{n\overline\sigma}(i) \end{array} \right).
\end{equation}
Here ${\mathcal{K}}^{\pm}_{ij,\sigma}=-t\delta_{\langle
ij\rangle}+(h_i-\mu)\delta_{ij}\pm\sigma m_i\delta_{ij}$, where
$\sigma=\pm1$ for up and down spin, $\delta_{ij}$ and
$\delta_{\langle ij \rangle}$ are the Kronecker delta symbol
connecting onsite and nearest neighbor sites, respectively. The
off-diagonal block ${\mathcal{D}}_{ij,\sigma}$ connects the
nearest neighbor links
${\mathcal{D}}_{ij,\sigma}=-\Delta_{ij}\delta_{ \langle ij
\rangle}$ with minus (plus) signs on the $a$($b$)-links for the
$d_{x^2-y^2}$-wave pairing symmetry, or on-site coupling
${\mathcal{D}}_{ij,\sigma}=-\Delta_i\delta_{ij}$ for conventional
$s$-wave pairing. We note that a modified Bogoliubov
transformation ${\hat{c}_{i\sigma}}^\dagger = \sum_{n}
(u^*_{n\sigma}(i) \hat{\gamma}_{n\sigma}^\dagger + v_{n\sigma}(i)
{\hat{\gamma}}_{n\overline{\sigma}})$, implemented in Ref.
\onlinecite{bobkova05}, led to modified amplitudes
$v_{n\sigma}(i)$: $v_{n\sigma}(i)\to{\sigma}v_{n\sigma}(i)$. The
corresponding basic equations for these modified amplitudes
coincide with those in the present paper (in particular, see
Eqs.(\ref{BdG}), (\ref{BdG1_ky_100})-(\ref{BdG2_ky_110}),
(\ref{Andr1}), (\ref{Andr2})) after redefining the off-diagonal
blocks
${\mathcal{D}}_{ij,\sigma}\to-\sigma{\mathcal{D}}_{ij,\sigma}$,
or, equivalently, $\Delta_{ij}\to-\sigma\Delta_{ij}$,
$\Delta_{i}\to-\sigma\Delta_{i}$.

\begin{figure}[!tbh]
\begin{minipage}{8.5cm}
  \begin{overpic}[scale=1]{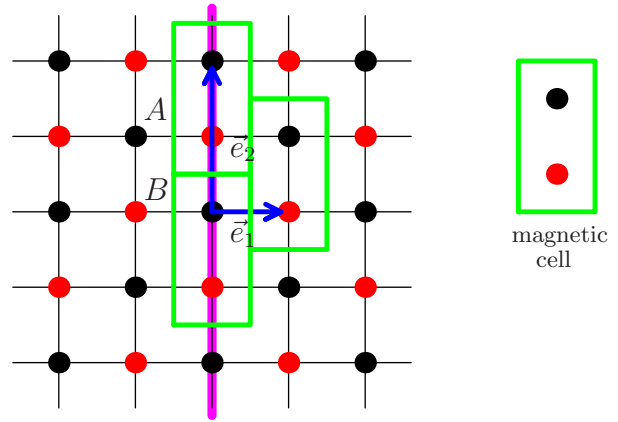}
    \put(24,50){\large $A$}
    \put(24,37){\large $B$}
    \put(38,30){\large $\vec e_1$}
    \put(38,44){\large $\vec e_2$}
    \put(84,30){$\rm magnetic$}
    \put(88,26){$\rm cell$}
            \end{overpic}
\end{minipage}
\caption{(Color online) $\{ 100 \}$ interface, showing the
corresponding unit cells with two atoms, and basis vectors $\vec
e_1$, $\vec e_2$ of the magnetic lattice.} \label{100}
\end{figure}

\begin{figure}[!tbh]
\begin{minipage}{8.5cm}
  \begin{overpic}[scale=1]{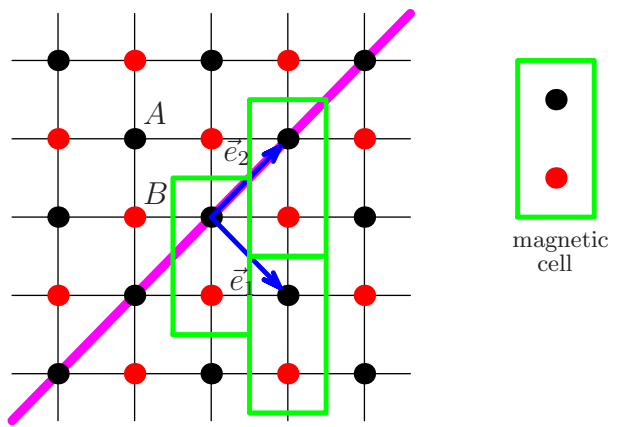}
    \put(24,50){\large $A$}
    \put(24,37){\large $B$}
    \put(38,23){\large $\vec e_1$}
    \put(37,44){\large $\vec e_2$}
    \put(84,30){$\rm magnetic$}
    \put(88,26){$\rm cell$}
            \end{overpic}
\end{minipage}
\caption{(Color online) $(110)$ interface, showing the
corresponding unit cells and basis vectors.} \label{110}
\end{figure}

Crystal periodicity along the interface makes it convenient to
Fourier transform the BdG equations along the $y$ axis and
introduce a wave vector component $k_y$. In the presence of
antiferromagnetic ordering, this should be done by taking into
account magnetic crystal symmetry along the boundary. In
particular, the magnetic order parameter $m_j$ oscillates rapidly
along the (100) interface and results in a doubling of the period
along the $y$-axis. In general, a modified magnetic period along
the boundary arises for all interface orientations except for a
(110) interface. We will take into account the magnetic crystal
symmetry by introducing a unit cell which contains two neighboring
atoms which belong to different magnetic sublattices $A$ and $B$.
Unit cells chosen below for the (100) and (110) interface are shown
in Figs. \ref{100} and \ref{110}, respectively.

On account of the magnetic crystal symmetry, the Fourier
transformation is taken to be of the form
\begin{equation}
\left(\begin{array}{c}
u_{{\bf j},\sigma}^{A(B)}\\
\\
v_{{\bf j},\bar\sigma}^{A(B)}\\
\end{array}\right)=\dfrac{d_y}{2\pi}\int_{-\pi/d_y}^{\pi/d_y}dk_y
e^{ik_yd_yj_y}
\left(\begin{array}{c}
u_{j_x,\sigma}^{A(B)}(k_y)\\
\\
v_{j_x,\bar\sigma}^{A(B)}(k_y)\\
\end{array}\right).
\label{fourier}
\end{equation}
Here $k_y$ is measured in units of $a^{-1}$ and $d_y=2,\,
\sqrt{2}$ for $(100),\, (110)$ interfaces, respectively. The
transformation is identical for atoms of sublattices $A$ and $B$
in the same unit cell $\bm j$. The vector ${\bm j}=(j_x,j_y)$
denotes cell coordinates, where $j_{x(y)}$ is the $x(y)$
coordinate of the cell measured in units of the appropriate basis
vectors. For definiteness, we identify cell positions with
positions of the associated site $A$.

Let vector $\bm i$ denote the location of a nearest neighbor site
for site $A(B)$ in the same unit cell. Then, the positions of all
nearest neighbors are described in the case of a (100) interface
in terms of the basis vectors shown in Fig. \ref{100} as $\langle
{\bm i}+{\bm e}_1 \pm \frac{\displaystyle {\bm e}_2}{\displaystyle
2}, {\bm i}-{\bm e}_1 \pm \frac{\displaystyle {\bm
e}_2}{\displaystyle 2}, {\bm i} \pm {\bm e}_2, {\bm i} \rangle$.
Taking this into account when performing the Fourier
transformation (\ref{fourier}) in Eqs.(\ref{BdG}), we obtain the
following one-dimensional Bogoliubov-de Gennes equations for the
(100) case:
\begin{widetext}
\begin{multline}
-\mu u_{j,\sigma}^{A(B)}(k_y)-t e^{\pm i k_y}\left(
u_{j+1,\sigma}^{B(A)}(k_y) + u_{j-1,\sigma}^{B(A)}(k_y) + 2 \cos k_y
u_{j,\sigma}^{B(A)}(k_y) \right)+\sigma m_j^{A(B)}u_{j,\sigma}^{A(B)}(k_y)
-\Delta_{s,j}^{A(B)}v_{j \bar \sigma}^{A(B)}(k_y)\\
-e^{\pm i k_y}\left(\Delta_{d,jj+1}^{A(B),a}v_{j+1,\bar
\sigma}^{B(A)}(k_y) + \Delta_{d,jj-1}^{A(B),a}v_{j-1,\bar\sigma}^{B(A)}(k_y)
+ 2 \cos k_y \Delta_{d,jj}^{A(B),b}v_{j,\bar \sigma}^{B(A)}(k_y) \right)
=E u_{j,\sigma}^{A(B)}(k_y)
\enspace , \label{BdG1_ky_100}
\end{multline}
\begin{multline}
\mu v_{j,\bar \sigma}^{A(B)}(k_y)+t e^{\pm i k_y}\left( v_{j+1,\bar
\sigma}^{B(A)}(k_y) + v_{j-1,\bar \sigma}^{B(A)}(k_y) + 2 \cos k_y
v_{j,\bar \sigma}^{B(A)}(k_y) \right)+
\sigma m_j^{A(B)}v_{j,\bar \sigma}^{A(B)}(k_y)
- \Delta_{s,j}^{A(B)^*}u_{j,\sigma}^{A(B)}(k_y)\\
-  e^{\pm i k_y} \left(\Delta_{d,jj+1}^{A(B),\, a^*}u_{j+1,\sigma}^{B(A)}(k_y)
+ \Delta_{d,jj-1}^{A(B),\, a^*}u_{j-1,\sigma}^{B(A)}(k_y) +
2 \cos k_y \Delta_{d,jj}^{A(B),\, b^*}u_{j,\sigma}^{B(A)}(k_y)
\right)=E v_{j,\bar \sigma}^{A(B)}(k_y)
\label{BdG2_ky_100} \enspace .
\end{multline}
\end{widetext}
Here, $j$ denotes only the $x$ coordinate of a cell. The factors
$\exp(\pm ik_y)$ arise in the nonlocal terms in
Eqs.(\ref{BdG1_ky_100}), (\ref{BdG2_ky_100}) since for the (100)
interface orientation, in accordance with the definitions above,
the $y$ coordinate of the site $B$ is always less by $a$ than the
coordinate of site $A$ in the same cell.

For the (110) orientation, however, the $y$ coordinate of the site
$B$ is less by $a/\sqrt{2}$ than the $y$ coordinate of site $A$ in
the same cell. Furthermore, if the vector $\bm i$ denotes the
location of a nearest neighbor site for site $A(B)$ in the same
unit cell, the positions of all nearest neighbors in the case of
the (110) interface are described in terms of basis vectors shown
in Fig. \ref{110} as $\langle {\bm i}\mp{\bm e}_1, {\bm i}\pm{\bm
e}_2,{\bm i},{\bm i}\mp{\bm e}_1\pm{\bm e}_2 \rangle$. Thus, for
the $(110)$ interface orientation we obtain the following
one-dimensional BdG equations:
\begin{widetext}
\begin{multline}
-\mu u_{j,\sigma}^{A(B)}(k_y)-2 t \cos \frac{k_y}{\sqrt 2} e^{\pm
\frac{\displaystyle i k_y}{\displaystyle \sqrt 2}}\left(
u_{j,\sigma}^{B(A)}(k_y) + u_{j\mp 1,\sigma}^{B(A)}(k_y)\right)+
\sigma m_j^{A(B)}u_{j,\sigma}^{A(B)}(k_y) - \Delta_{s,j}^{A(B)}
v_{j,\bar\sigma}^{A(B)}(k_y)\\
-\left[\left(\Delta_{d,jj}^{A(B),a}e^{\displaystyle\pm\sqrt{2}i k_y}
+\Delta_{d,jj}^{A(B),b}\right)v_{j,\bar\sigma}^{B(A)}(k_y)
+\left(\Delta_{d,jj\mp1}^{A(B),b}e^{\displaystyle\pm\sqrt{2} i k_y}
+\Delta_{d,jj\mp1}^{A(B),a}\right) v_{j\mp 1,\bar\sigma}^{B(A)}(k_y)
\right]=E u_{j,\sigma}^{A(B)}(k_y)
\label{BdG1_ky_110} \enspace ,
\end{multline}
\begin{multline}
\mu v_{j,\bar\sigma}^{A(B)}(k_y)+2 t \cos \frac{k_y}{\sqrt 2} e^{\pm
\frac{\displaystyle i k_y}{\displaystyle \sqrt 2}}\left( v_{j,\bar
\sigma}^{B(A)}(k_y) + v_{j\mp 1,\bar\sigma}^{B(A)}(k_y)\right)+
\sigma m_j^{A(B)}v_{j,\bar\sigma}^{A(B)}(k_y) - \Delta_{s,j}^{A(B)^*}
u_{j,\sigma}^{A(B)}(k_y)\\
- \left[\left(\Delta_{d,jj}^{A(B),\, a^*}e^{\displaystyle \pm\sqrt{2}i k_y}
+\Delta_{d,jj}^{A(B),\, b^*}\right)u_{j,\sigma}^{B(A)}(k_y) +
\left(\Delta_{d,jj\mp1}^{A(B),\, b^*}e^{\displaystyle\pm \sqrt{2}i k_y}
+\Delta_{d,jj\mp1}^{A(B),\, a^*}\right)u_{j\mp1,\sigma}^{B(A)}(k_y) \right] =
E v_{j,\bar\sigma}^{A(B)}(k_y)
\label{BdG2_ky_110} \enspace .
\end{multline}
\end{widetext}

The singlet superconducting order parameters entering
Eqs.(\ref{BdG1_ky_100})-(\ref{BdG2_ky_110}), are defined as
$\Delta^{A(B)}_{ij}=-(V_i/2)\langle
\hat{c}^{A(B)}_{i\downarrow}\hat{c}^{B(A)}_{j\uparrow} -
\hat{c}^{A(B)}_{i\uparrow}\hat{c}^{B(A)}_{j\downarrow} \rangle$.
The magnetic order parameter is $m^{A(B)}_i=(U_i/2)\left[ \langle
\hat{n}^{A(B)}_{i\uparrow} \rangle\ - \langle
\hat{n}^{A(B)}_{i\downarrow}\rangle\right]$. For the study of
proximity effects, it is convenient to introduce the magnetization
$M_i$ and the pairing amplitude $F_{ij}$, which are related to
$m_i$ and $\Delta_{ij}$ by $m_i=U_i M_i$ and $\Delta_{ij}=-V_i
F_{ij}$, respectively.

The self-consistency equations in the sublattice representation
take the form
\begin{widetext}
\begin{equation}
{n}^{A(B)}_{i\sigma}=\sum_{n,k_y}\left[
|u_{n,i_x,\sigma}^{A(B)}(k_y)|^2f(E_{n,k_y,\sigma})+
|v_{n,i_x,\sigma}^{A(B)}(k_y)|^2 f(-E_{n,k_y,\sigma})\right],
\end{equation}

\begin{eqnarray}\nonumber
\Delta^{A(B)}_{ij}=-\frac{V_i}{2}\sum_{n,k_y,\sigma}\left[
u_{n,i_x,\sigma}^{A(B)}(k_y)v^{B(A)^*}_{n,j_x,\bar\sigma}(k_y)e^{\displaystyle
ik_yd_y(i_y-j_y)}f(-E_{n,k_y,\sigma})-\right.\\ \left.
u_{n,j_x,\sigma}^{B(A)}(k_y)v^{A(B)^*}_{n,i_x,\bar\sigma}(k_y)e^{\displaystyle
-ik_yd_y(i_y-j_y)}f(E_{n,k_y,\sigma})\right]\, . \label{selfd}
\end{eqnarray}
\end{widetext}
The sum is taken over eigenstates of
Eqs.(\ref{BdG1_ky_100})-(\ref{BdG2_ky_110}), which depend on
$k_y$, $\sigma$ and possibly an additional set of quantum numbers
${n}$. Eq.(\ref{selfd}) applies to the $d$-wave case, whereas for
the $s$-wave superconductor with on-site pairing one should put in
Eq.(\ref{selfd}) $\Delta^{\alpha}_{ii}$ with amplitudes
$u^\alpha_{n,i_x,\sigma}(k_y) $,
$v^\alpha_{n,i_x,\bar\sigma}(k_y)$ taken for one sublattice
$\alpha=A,B$. As usual, $f(E)=[1+\exp(E/T)]^{-1}$ denotes the
Fermi distribution function at temperature $T$.

Obviously, any bond between nearest neighbors connects two sites
from different sublattices. The notation $\Delta^{A}_{ij}$
($\Delta^{B}_{ij}$) means that the order parameter is taken on the
bond which connects a site of sublattice $A$ ($B$) within the unit
cell $i$ with a nearest neighbor site (of sublattice $B$ ($A$)) in
the unit cell $j$.  All order parameters are presumed to be
identical on links (or sites) which can be obtained from each
other by magnetic translations along the interface or interlayer.
For this reason, it is sufficient to consider $i$ and $j$ in the
notation $\Delta^{A(B)}_{ij}$ as containing only $x$ components,
if one indicates in addition the type of link ($a$ or $b$).

\section{Andreev equations and ${\cal S}$-matrices}

We will base our numerical calculations on the one-dimensional
Bogoliubov-de Gennes equations formulated above, as well as on the
corresponding self-consistency equations. For an analytical study
of superconducting interfaces and junctions involving itinerant
antiferromagnets, we present in this section the quasiclassical
approach developed in Ref. \onlinecite{bobkova05}. As is well
known, the quasiclassical theory describes physical quantities
which vary slowly in space compared to the atomic scale, and
assumes characteristic energies to be much less than the Fermi
energy $E_F$. We consider below two types of
superconductor-antiferromagnet hybrid systems to which the
quasiclassical approach applies to a different extent.

The first type of system satisfies the conditions $|\Delta|\ll
E_F$, $|m|\ll E_F$. The latter inequality guarantees that the
antiferromagnetic order parameter $m$, as a rule, varies slowly
within each separate sublattice $A$ and $B$. Hence, one can use
quasiclassical equations both for superconducting and
antiferromagnetic phases and match them at the interface if they
are formulated separately for each sublattice. Sublattice
equations are coupled with each other via nonlocal terms which
contain, for example, hopping matrix elements or $d$-wave
superconducting order parameter fields. In total, this gives us
twice the usual number of coupled quasiclassical equations.
Another possible formulation of the quasiclassical approach to the
first type of superconductor - itinerant antiferromagnet hybrid
systems is not based on the sublattice representation. Instead,
one can formulate equations for quasiparticle trajectories, taking
into account that the rapidly oscillating antiferromagnetic order
parameter $m_j=(-1)^{j_a+j_b}m =\exp(i\bm Q \bm j)m$ couples
equations for two trajectories, one for a quasiparticle momentum
$\bm k_F$ and the other for $({\bm k}_F+\bm Q)$. The approach
based on this $(\bm{k}_F,\bm{k}_F+\bm{Q})$ representation results
also in twice the number of quasiclassical equations. There are no
further coupled trajectories, since $2\bf Q$ is assumed to be the
reciprocal vector of the nonmagnetic crystal due to the nesting
condition.

In the second type of hybrid superconductor-antiferromagnet
systems, only the superconducting order parameter satisfies the
condition $|\Delta|\ll E_F$ and can be safely described with the
quasiclassical equations. The antiferromagnetic order parameter is
taken sufficiently large $\Delta\ll |m|\lesssim E_F$ (one could
also assume $m\gtrsim E_F$ within the framework of the approach,
if this were relevant). Then the effect of the antiferromagnet on
the superconductor can be taken into account entirely via modified
boundary conditions, which complement quasiclassical equations at
abrupt superconductor - antiferromagnetic interfaces.

\subsection{Andreev equations in the sublattice representation}
We begin with the derivation of Andreev equations in the
sublattice representation. Assume for this purpose that the
solution of the BdG equations
(\ref{BdG1_ky_100})-(\ref{BdG2_ky_100}) and
(\ref{BdG1_ky_110})-(\ref{BdG2_ky_110}) can be represented as the
following product of rapidly oscillating exponents and a slowly
varying Andreev amplitude:
\begin{equation}
\left(
\begin{array}{c}
\vspace{0.1cm} u_{j\sigma}^A \\
\vspace{0.1cm} u_{j\sigma}^B \\
\vspace{0.1cm} v_{j\bar \sigma}^A \\
\vspace{0.1cm} v_{j\bar \sigma}^B \\
\end{array}
\right) = \exp\left({\displaystyle i \frac{{\bm k}_F \hat{\bm b}}{2}
\hat \gamma_3}\right)
\left(\begin{array}{c}
\vspace{0.1cm} \tilde u_{j\sigma}^A \\
\vspace{0.1cm} \tilde u_{j\sigma}^B \\
\vspace{0.1cm} \tilde v_{j\bar \sigma}^A \\
\vspace{0.1cm} \tilde v_{j\bar \sigma}^B  \\
\end{array}\right)
\exp({\displaystyle i k_{F,x} d_x j}) \label{amp_100_110} \enspace .
\end{equation}
Here, $k_{F,x}$ is the $x$ component of the quasiparticle momentum
on the Fermi surface, measured in units of $a^{-1}$. The quantity
$k_{F,x}$ can be considered a function of $k_y$. The Pauli matrix
$\hat \gamma_3$ operates in $\{AB\}$ sublattice space, $d_x=1$ for
$(100)$ interface and $d_x=\sqrt 2$ for the $(110)$ case.
Introducing the unit vector $\hat {\bf b}$ along the crystal
$b$-axis permits us to define the Andreev amplitudes in
Eq.(\ref{amp_100_110}) in a unified form which applies to all
interface orientations.

\begin{figure}[!tbh]
\includegraphics[width=7cm,height=5cm]{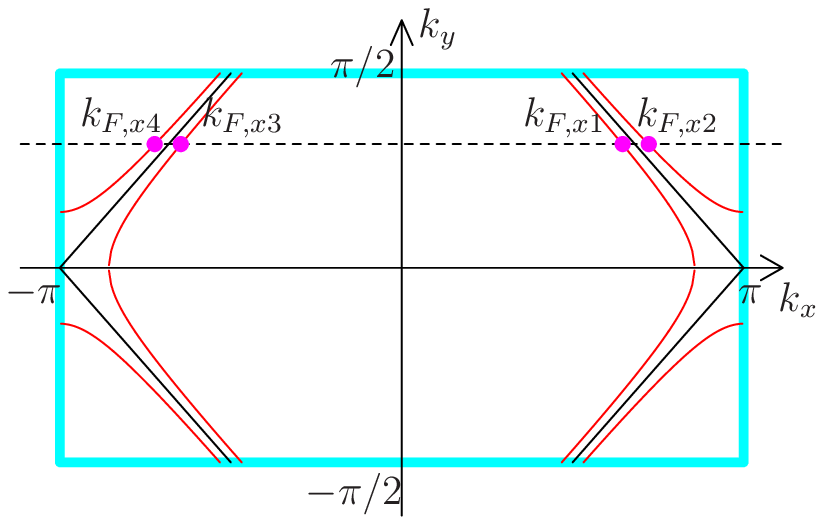}
\caption{(Color online) The Brillouin zone adapted to the (100)
interface and the Fermi surfaces for small and vanishing $\mu$.}
\label{zb_fs_100}
\end{figure}

\begin{figure}[!tbh]
\includegraphics[width=8.5cm, height=5cm]{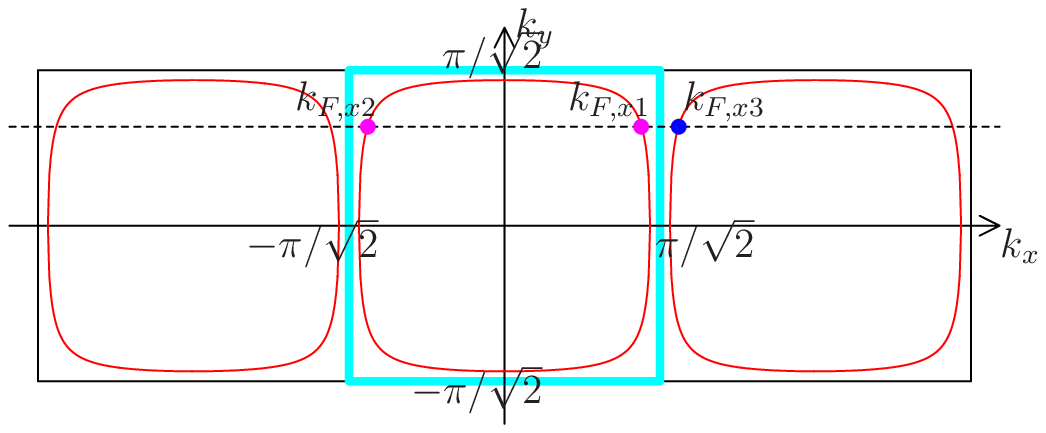}
\caption{(Color online) The Brillouin zone adapted to (110)
interface and the Fermi surfaces for small and vanishing $\mu$.}
\label{zb_fs_110}
\end{figure}

As mentioned above, the parameter $\mu$ is considered throughout
the paper to be small $\mu\ll E_F$, since the antiferromagnetic
phase is stable only close to half-filling. For this reason, one
can additionally include effects of small deviations from
half-filling in the quasiclassical approximation. Taking this into
account, we define a small parameter of the quasiclassical
expansion as $\alpha={\rm max}(|m|, |\Delta|, |\mu|)a/|v_{F,x}|\ll
1$. Here $v_{F,x}$ is the $x$-component of the Fermi velocity in
the normal metal state at half filling. The quasiclassical
approach works only for those $k_y$ for which $v_{F,x}$ is not too
small and does not break the condition $\alpha\ll 1$. We expand
all properties associated with the Fermi surface in powers of the
small parameter $(\mu a/|v_{F,x}|)\ll 1$. This concerns, in
particular, $k_{F,x}$ which enters the exponential factor in
Eq.(\ref{amp_100_110}). The surface adapted Brillouin zone and
respective Fermi surface for small and vanishing $\mu$ are shown
in Figs.\ref{zb_fs_100} and \ref{zb_fs_110} for the (100) and
(110) interface orientations, respectively.

For the (100) orientation, the normal-state electron band is
described as $\xi^{\pm}({\bf k})= \mp 2 t (\cos k_x + \cos k_y) -
\mu$, and the Brillouin zone in the case of two atoms in unit cell
(see Fig. \ref{100}) is spanned by $k_{x} \in [-\pi,\pi]$, $k_{y}
\in [-\pi/2,\pi/2]$. For the half-filled band $k_{F,x}=\pm
(\pi-|k_y|)$ and $v^{\pm}_{F,x}=\pm 2ta\sin k_{F,x}$. As is seen
in Fig. \ref{zb_fs_100}, four possible values of $k_{F,x}$, which
occur for a given $k_y$ at $\mu\ne 0$, merge into two values at
$\mu=0$.

For the (110) interface, we have in the sublattice representation
$\xi^{\pm}({\bm k})=\mp 4t\cos\bigl(k_x/\sqrt{2}\bigr)\cos\bigl(
k_y/\sqrt{2} \bigr) - \mu$ and
$k_{x,y}\in[-\pi/\sqrt{2},\pi/\sqrt{2} ]$. If $\mu>0$, only the
Fermi surface $\xi^{-}({\bm k})=0$ exists in the first Brillouin
zone, whereas for $\mu<0$ the Fermi surface is determined from
$\xi^{+}({\bm k})=0$. At $\mu=0$ both Fermi surfaces coincide with
the edges of the first Brillouin zone. As seen in Fig.
\ref{zb_fs_110}, two different values of $k_{F,x}$, which occur
within the first Brillouin zone for a given $k_y$, touch the edges
of the zone $k_{F,x}=\pm\pi/ \sqrt{2}$ in the case of half filling
and hence, become equivalent at $\mu=0$. Since the Fermi
velocities $v_{F,x_1}$ and $v_{F,x_3}$ have opposite signs, for
the half-filled band $v^{\pm}_{F,x}(\bm {k}_F)=\pm
2\sqrt{2}ta\cos\left( k_y/\sqrt{2}\right )$, where two signs of
$v_{F,x}(\bm {k}_F)$ at the same $\bm {k}_F$ correspond to two
degenerate parts of the Fermi surface.

To obtain the Andreev equations for the (100) orientation, we
substitute the ansatz (\ref{amp_100_110}) into
Eqs.(\ref{BdG1_ky_100}), (\ref{BdG2_ky_100}), separately for each
of the two values $k_{F,x}=\pm (\pi-k_y)$, and disregard terms
$\lesssim\alpha^2$. For the (110) interface one should proceed
analogously, using in Eqs.(\ref{BdG1_ky_110}), (\ref{BdG2_ky_110})
the ansatz (\ref{amp_100_110}) with $k_{F,x}=\pi/\sqrt{2}$.  We
note that $|\tilde{u}_{j\pm1,\sigma}^{B(A)} -
\tilde{u}_{j,\sigma}^{B(A)}|\lesssim
\alpha|\tilde{u}_{j,\sigma}^{B(A)}|$,
$|\tilde{u}_{j+1,\sigma}^{B(A)}+ \tilde{u}_{j-1,\sigma}^{B(A)} -
2\tilde{u}_{j,\sigma}^{B(A)}|\lesssim
\alpha^2|\tilde{u}_{j,\sigma}^{B(A)}|$. Neglecting terms of the
order of $\alpha^2$, one gets with the required accuracy $\tilde
u_{j + 1}^\nu - \tilde u_j^\nu = \tilde u_j^\nu - \tilde u_{j -
1}^\nu = \left(\tilde u_{j + 1}^\nu - \tilde u_{j -
1}^\nu\right)/2 = d_x\partial \tilde u_j^\nu /\partial x_j$.

Using the outlined procedure, we obtain the following Andreev equations
in the sublattice representation
\begin{widetext}
\begin{eqnarray}
- \mu \tilde u_{j,\sigma}^{A(B)} - i v^+_{F,x}\frac{\partial \tilde
u_{j,\sigma}^{B(A)}}{\partial x_j} + \sigma m_j^{A(B)} \tilde
u_{j,\sigma}^{A(B)} - \Delta_{s,j}^{A(B)} \tilde v_{j,
\bar\sigma}^{A(B)}
- \Delta_{d,j}^{A(B)}(k_{F,x},k_y)\tilde v_{j , \bar\sigma}^{B(A)}=
E_\sigma \tilde u_{j,\sigma}^{A(B)} \label{Andr1}
\enspace ,       \\
\mu \tilde v_{j , \bar\sigma}^{A(B)} + i v^+_{F,x}\frac{\partial
\tilde v_{j , \bar\sigma}^{B(A)}}{\partial x_j} + \sigma m_j^{A(B)}
\tilde v_{j , \bar\sigma}^{A(B)} - \Delta_{s,j}^{A(B)^*}
\tilde u_{j,\sigma}^{A(B)}
- \Delta_{d,j}^{A(B)^*}(k_{F,x},k_y)\tilde u_{j,\sigma}^{B(A)}=
E_\sigma \tilde v_{j , \bar\sigma}^{A(B)}
\label{Andr2} \enspace .
\end{eqnarray}
\end{widetext}
These equations take a unified form, which applies to any
interface orientation. For the (100) orientation
$\Delta_{d,j}^{A(B)}(k_{F,x},k_y)=2\Delta_{d,j}^{A(B)}(\cos
k_{F,x}-\cos k_y)$. The on-site $d$-wave order parameter
$\Delta_{d,j}^{A(B)}$, slowly varying on the atomic scale with
coordinate $j$ along $x$ axis, is defined in the coordinate space
by the four surrounding links. With standard site coordinates
$\bm{i} =(i_x,i_y)$ one can write
$\Delta_{d,i}\!=\!\frac{1}{4}\left(\Delta_{d,
ii+a}\!+\!\Delta_{d,ii-a}\!-\!\Delta_{d,ii+b}\!-\!\Delta_{d,ii-b}\right)$.
For the (110) orientation we have
$\Delta_{d,j}^{A(B)}(k_{F,x},k_y)\!=\!-4\Delta_{d,j}^{A(B)}
\sin(k_{F,x}/\sqrt 2)\sin({k_y}/{\sqrt 2})$ with
$k_{F,x}\!=\!\pi/\sqrt{2}$.

As one can see from the derivation of Eqs.(\ref{Andr1}),
(\ref{Andr2}), they apply to various cases when the sublattice
magnetic and/or superconducting order parameters vary slowly in
space, satisfying standard quasiclassical conditions. For
instance, no particular relation between the sublattice
magnetizations $m^A_j$ and $m^B_j$ is implied yet. For both
interface orientations, the Fermi velocity $v^+_{F,x}(\bm{k}_F)$
is positive for $k_{F,x}>0$ within the Brillouin zone in the
sublattice representation. However, the associated solutions of
Eqs.(\ref{Andr1}), (\ref{Andr2}) can describe, in general, both
incoming and outgoing quasiparticles on either side of the
interface. This is seen from the expression for the density of a
quasiparticle probability current $\bm{j}_P$, which can be found
from the BdG equations in much the same standard way known in the
case of the Schr\"odinger equation. The probability current
density along the $x$ axis, carried by the solution with quantum
numbers ($n,\sigma,k_y$), can be written in the $(100)$ case as
\begin{equation}
j_{P,x} = \frac{v^+_{F,x}}{2 a}\sum \limits_{\alpha = \pm,\nu}
\left\{ \alpha (\tilde u_{n,\sigma,\alpha}^{j, \nu *} \tilde
u_{n,\sigma,\alpha}^{j,\bar \nu} - \tilde v_{n,\sigma,\alpha}^{j,
\nu} \tilde v_{n,\sigma,\alpha}^{j,\bar \nu *}) \right\}
\label{Jfull_100} \enspace .
\end{equation}
Here, the sum is taken over sublattice index $\nu=A, B$, as well
as over the two parts of the Fermi surface $\alpha=\pm 1$.

For the (110) orientation we find
\begin{equation}
j_{P,x} = \frac{v^+_{F,x}}{2 a}\sum \limits_{\nu} \left\{(\tilde
u_{n,\sigma}^{j, \nu *} \tilde u_{n,\sigma}^{j,\bar \nu} - \tilde
v_{n,\sigma}^{j, \nu} \tilde v_{n,\sigma}^{j,\bar \nu *}) \right\}
\label{Jfull_110} \enspace .
\end{equation}
As usual, the components $u$ and $v$ of the Andreev amplitudes
with the same wave vector have opposite contributions to the
probability current\cite{Andreev64}. Since the current is formed
mainly by hopping between nearest neighbor sites, it is determined
in the sublattice representation as a mixed product of $A$ and $B$
components of Andreev amplitudes for any interface orientation.
Hence, the sign of $j_{P,x}$ depends not only on the crystal wave
vector $\bm{k}_{F}$, but also on the {\it relative} signs of the
$A$ and $B$ components of the Andreev amplitudes.

One can further transform the equations (\ref{Andr1}),
(\ref{Andr2}), which are formulated in the $\{AB\}$-sublattice
representation with two atoms per unit cell into the
representation with one atom per unit cell. Consider, for example,
the $(110)$ interface orientation. For one atom per unit cell, the
Brillouin zone is spanned by $k_{x}\in[-\sqrt{2}\pi,
\sqrt{2}\pi]$, $k_{y}\in [-\pi/\sqrt{2},\pi/\sqrt{2}]$ and
$\xi({\bm k})$ is given in Eq.(\ref{xi1_110}), whereas in the case
of two atoms per unit cell the Brillouin zone in the $k_x$
direction is: $k_{x}\in [-\pi/\sqrt{2},\pi/\sqrt{2}]$. Assuming
$|k_{F,x}|\le\pi/\sqrt{2}$, one can write the following relation
between the quasiparticle amplitudes in the two representations
(compare with Eq.(\ref{amp_100_110})):
\begin{widetext}
\begin{eqnarray}\nonumber
{\tilde u_{j}^A (\bm{k}_{F})e^{i\left(k_y-k_{F,x}\right)/2\sqrt{2}}\choose
\tilde u_{j}^B (\bm{k}_F)e^{-i\left(k_y-k_{F,x}\right)/2\sqrt{2}}}
e^{i k_{F,x}\sqrt{2}j}={\tilde u_{2j}(\bm{k}_{F}) \choose
\tilde{u}_{2j+1}(\bm {k}_{F})e^{-i\left(k_y-k_{F,x}\right)/\sqrt{2}}}
e^{i (k_{F,x}/\sqrt{2})2j}\\
+{\tilde u_{2j}(\bm{k}_F+\bm{Q})\choose -\tilde u_{2j+1}(\bm{k}_F+\bm{Q})
e^{-i\left(k_y-k_{F,x}\right)/\sqrt{2}}}e^{i (k_{F,x}/\sqrt{2})2j}
\label{and_AB} \enspace .
\end{eqnarray}
\end{widetext}
Here, we have taken into account that if $j$ is the $x$ coordinate
of a two-atom unit cell, then in the representation with one atom
per unit cell, the site $A$ has even $x$ coordinate $2j$, whereas
site $B$ has odd coordinate $2j+1$.

In the sublattice representation, the wave vector $\bm
Q=(\pm\sqrt{2} \pi,0)$, which is the wave vector of the
antiferromagnetic pattern that we will study below, is the
reciprocal crystal vector. Thus, the wave vectors $\bm{k}_{F}$ and
$\bm{k}_{F}+\bm Q$ are equivalent to each other in the approach
with two atoms per unit cell. In the representation with one atom
per unit cell, the wave vectors $\bm{k}_{F}$ and $\bm{k}_{F}+\bm
Q$ are physically different. The quantity $\bm{k}_{F}+\bm Q$ in
Eq.(\ref{and_AB}) is assumed to lie in the first Brillouin zone of
the representation with one atom per unit cell, so that $\bm{Q}=
(\pm\sqrt{2}\pi,0)$ should be taken there with minus sign for $0<
k_{F,x}\le\pi/\sqrt{2}$ and with plus sign for $-\pi/\sqrt{2}\le
k_{F,x}\le 0$. Thus, it follows from Eq.(\ref{and_AB}) that
\begin{equation}
{\tilde u_{j}^A (\bm{k}_F)\choose \tilde u_{j}^B (\bm{k}_F)}\!
=\!{\!\tilde u_{2j}(\bm{k}_{F})+\tilde u_{2j}(\bm{k}_F\!+\bm{Q})\! \choose
\!\tilde{u}_{2j+1}(\bm {k}_{F})-\tilde u_{2j+1}(\bm{k}_F\!+\bm{Q})\!}
e^{\!-i\frac{k_y-k_{F,x}}{2\sqrt{2}}}\, .
\label{AB_1}
\end{equation}

For the $(100)$ interface orientation, the Brillouin zone for the
square lattice with one atom per unit cell is spanned by $k_{x,y}
\in [-\sqrt{2}\pi,\sqrt{2}\pi]$ and $\xi({\bm k})$ is given in
Eq.(\ref{xi1_100}), whereas for two atoms per unit cell, shown in
Fig. \ref{110}, the Brillouin zone in the $k_y$ direction is:
$k_{y}\in [-\pi/\sqrt{2},\pi/\sqrt{2}]$. Assuming
$|k_{F,y}|\le\pi/\sqrt{2}$, we find the following relation between
the quasiparticle amplitudes in the two representations:
\begin{equation}
{\tilde{u}_j^A (k_y) \choose \tilde{u}_j^B (k_y)} =
\frac{1}{2}e^{-i k_y/2}{\tilde{u}_j(k_y)+\tilde u_j(k_y+Q_y)
\choose \tilde u_j(k_y)-\tilde u_j(k_y+Q_y)}
\label{AB_2} \enspace .
\end{equation}
Here, the wave vector $\bm Q=(\pm\pi,\pm \pi)$ is the reciprocal
crystal vector in the sublattice representation. The quantity
$k_y+Q_y$ in Eq.(\ref{AB_2}) is assumed to lie in the first
Brillouin zone of the representation with one atom per unit cell,
so that $Q_y=\pm\pi$ should be taken with minus sign for
$0<k_{F,y}\le\pi/\sqrt{2}$ and with plus sign for
$-\pi/\sqrt{2}\le k_{F,y}\le 0$.

Substituting Eq.(\ref{AB_1}) or (\ref{AB_2}) into Eqs.
(\ref{Andr1}), (\ref{Andr2}), we obtain the Andreev equations in
the $(\bm{k}_F,\bm{k}_F+ \bm{Q})$-representation. Since
quasiparticle crystal momenta $\bm{k}_{F}$ and $\bm{k}_{F}+\bm Q$
both lie on the Fermi surfaces on either side of the interface,
the Andreev equation in the $(\bm{k}_F,\bm{k}_F
+\bm{Q})$-representation takes the comparatively simple form:
\begin{widetext}
\begin{eqnarray}\nonumber
\Bigl( - i v_{F,x}(\bm k_{F}) \hat \rho_z \hat \tau_z
\frac{\partial}{\partial x_j} - \mu \hat \tau_z + \sigma
\frac{m_j^A+m_j^B}{2} + \sigma \frac{m_j^A-m_j^B}{2}\hat \rho_x -
\Delta_{s,j}\frac{\hat \tau_+}{2}-\\
\Delta_{s,j}^*\frac{\hat \tau_-}{2} -\Delta_{d,j}(\bm k_{F})
\frac{\hat \tau_+}{2} \hat \rho_z - \Delta_{d,j}^{*}(\bm k_{F})
\frac{\hat \tau_-}{2}\hat \rho_z \Bigr) \hat \Psi_{j \sigma} =
E_\sigma \hat \Psi_{j\sigma} \label{AndrkQ} \enspace .
\end{eqnarray}
\end{widetext}
Here, $\hat \Psi_{j \sigma} = (\tilde u_{j\sigma}(\bm k_F), \tilde
u_{j\sigma}(\bm k_F + \bm Q),\tilde v_{j{\bar\sigma}}(\bm
k_F),\tilde v_{j{\bar\sigma}}(\bm k_F+ \bm Q))$. $\hat\tau_\alpha$
and $\hat\rho_\alpha$ denote the Pauli matrices in particle-hole and
$\{\bm k_F, \bm k_F + \bm Q\}$ spaces. Eq.(\ref{AndrkQ}) applies
generally for any particular relation between $m_j^A$ and $m_j^B$.
For instance, it applies also to the study of weak ferromagnets
($m \ll \varepsilon_f$). In the case of antiferromagnetic ordering
satisfying the condition $m_j^B = - m_j^A $, within the quasiclassical
accuracy one can disregard in Eq.(\ref{AndrkQ}) the term containing
$m_j^A+m_j^B$.

The only term which couples Andreev amplitudes with momenta
$\bm{k}_F$ and $\bm{k}_F+\bm{Q}$ in Eq.(\ref{AndrkQ}) contains the
difference between sublattice magnetizations $m_j^A-m_j^B$. This
is natural, since a finite difference $m_j^A-m_j^B$ results in a
doubling of the period in the system which we study. Eqs.
(\ref{Andr1}), (\ref{Andr2}) and (\ref{AndrkQ}) apply also in the
absence of period doubling, being equivalent in this case to the
standard Andreev equations. If period doubling takes place only at
the boundaries, the sublattice or $(\bm k_F, \bm k_F + \bm Q)$
representations of the quasiclassical equations can be convenient
for applying appropriate boundary conditions to the solutions.

\subsection{Boundary conditions and {\cal S}-matrix in $(\bm k_F,
\bm k_F + \bm Q)$ representation}

The assumption of slowly varying order parameters does not apply
in the vicinity of abrupt boundaries. This invalidates
quasiclassical equations close to the boundaries and makes it
necessary to complement them with appropriate boundary conditions.
The conditions have been obtained for Andreev amplitudes in Ref.
\onlinecite{shelankov80} and re-derived later by various methods
(see, e.g. Ref. \onlinecite{shumeiko97}). The boundary conditions
for Andreev amplitudes at a plane interface can be written in the
following form
\begin{equation}
{ {\Psi}_-^l \choose {\Psi}_+^r} = \left(
\begin{array}{cc}
\check S_{11} & \check S_{12} \\
\check S_{21} & \check S_{22} \\
\end{array}
\right) { {\Psi}_+^l \choose {\Psi}_-^r}
\label{bc_100}\enspace .
\end{equation}
Here, $\Psi$ denotes a collection of Andreev amplitudes. For
example, in the $(\bm k_F, \bm k_F + \bm Q)$ representation it
contains eight amplitudes $\Psi^{T}_j=({\tilde u}_{j\sigma}(\bm
k_F), {\tilde u}_{j\sigma} (\bm k_F + \bm Q),{\tilde
v}_{j{\bar\sigma}}(\bm k_F),{\tilde v}_{j{\bar\sigma}}(\bm k_F+
\bm Q), {\tilde u}_{j\bar\sigma}( \bm{k}_F), {\tilde
u}_{j\bar\sigma}({\bm k}_F + \bm Q), {\tilde v}_{j{\sigma}}({\bm
k}_F),{\tilde v}_{j{\sigma}}(\bm{k}_F+ \bm{Q}))$. The superscripts
$l(r)$ indicate that the amplitudes are taken on the left (right)
side of the boundary. The subscripts $\pm$ in the amplitudes
denote the sign of the Fermi velocity components
$v_{F,x}(\bm{k}_{F})$ or $v_{F,x}(\bm{k}_{F}+\bm{Q})$ for
electrons. Thus, the solutions entering the left and right hand
sides of Eq.(\ref{bc_100}) are connected at the interface by the
normal-state scattering ${\cal S}$ matrix: ${\cal S}=\|\check{\cal
S}_{ij}\|$ ($i(j)=1,2$). This matrix $\check{\cal S}_{ii}$
contains the reflection amplitudes of normal-state quasiparticles
from the interface in $i$-th half-space, whereas $\check{\cal
S}_{ij}$ with $i\ne j$ incorporates the transmission amplitudes of
normal-state quasiparticles from side $j$.  In the $(\bm k_F, \bm
k_F + \bm Q)$ representation, each component $\check{\cal S}_{ij}$
($i(j)=1,2$) in the matrix $\cal S$ is an $8\times8$ matrix in the
eight-dimensional product space of particle-hole, spin and $(\bm
k_F, \bm k_F + \bm Q)$ variables. We introduce the Pauli matrices
$\rho_\alpha$, $\tau_\alpha$ and $\sigma_\alpha$, which operate in
$\left\{\bm k_F, \bm k_F + \bm Q\right\}$ space, particle-hole
space and spin space respectively.

The normal-state $\cal S$ matrix  is diagonal in particle-hole space:
\begin{equation}
\check S = \left(
\begin{array}{cc}
\check S_{11} & \check S_{12} \\
\check S_{21} & \check S_{22} \\
\end{array}
\right) = {\hat S} \frac{1+\tau_z}{2} + {\hat {\widetilde S}}
\frac{1-\tau_z}{2} \label{S_tau_100} \enspace .
\end{equation}

If the AF order parameter does not change its direction, one can
take a quantization axis along $\bm m$. In this case up and down
spin states are eigenstates of the BdG and Andreev equations,
which are formulated separately for each electron spin orientation
as given in Eqs.(\ref{BdG}),
(\ref{BdG1_ky_100})-(\ref{BdG2_ky_110}), (\ref{Andr1}), (\ref{Andr2}).
Then the associated $\cal S$ matrix turns out to be diagonal also
in spin space, and the boundary conditions reduce to the following
equalities:
\begin{multline}
\left\{\begin{array}{c}
{\tilde u}^{\alpha,l}_{\sigma,-}=\sum\limits_{\beta}\left(
S_{11,\sigma}^{\alpha\beta}{\tilde u}^{\beta,l}_{\sigma,+}+
S_{12,\sigma}^{\alpha\beta}{\tilde u}^{\beta,r}_{\sigma,-}\right)\\
\\
{\tilde u}^{\alpha,r}_{\sigma,+}=\sum\limits_{\beta}\left(
S_{21,\sigma}^{\alpha\beta}{\tilde u}^{\beta,l}_{\sigma,+}+
S_{22,\sigma}^{\alpha\beta}{\tilde u}^{\beta,r}_{\sigma,-}\right)
\enspace,
\end{array}
\right.
\label{bcu}
\end{multline}

\begin{multline}
\left\{\begin{array}{c}
{\tilde v}^{\alpha,l}_{\sigma,-}=\sum\limits_{\beta}\left(
{\widetilde S}_{11,\sigma}^{\alpha\beta}{\tilde v}^{\beta,l}_{\sigma,+}+
{\widetilde S}_{12,\sigma}^{\alpha\beta}{\tilde v}^{\beta,r}_{\sigma,-}\right)\\
\\
{\tilde v}^{\alpha,r}_{\sigma,+}=\sum\limits_{\beta}\left(
{\widetilde S}_{21,\sigma}^{\alpha\beta}{\tilde v}^{\beta,l}_{\sigma,+}+
{\widetilde S}_{22,\sigma}^{\alpha\beta}{\tilde v}^{\beta,r}_{\sigma,-}\right)
\enspace.
\end{array}
\right.
\label{bcv}
\end{multline}
Here, the superscripts $\alpha$, $\beta$ take the two values
$k_{F,y}$, $k_{F,y} + Q_y$.

As an example, we apply in the following the quasiclassical
approach to analytical calculations of the subgap spectrum of
quasiparticle interface states near a $(110)$ SC/AF interface in
the absence of potential barriers. A detailed self-consistent
study of quasiparticle states at interfaces with antiferromagnets
will be presented below in Sec. IV and V. Let an $s$-wave
superconductor and an antiferromagnet, separated with the (110)
interface, occupy separately the right ($j>0$) and the left
($j\le0$) half-spaces of the square lattice. Assuming
$|E|<|\Delta|\ll |m|, |v_{F,x}|/a$, we use the Andreev equations
only in the superconducting region and apply the boundary
conditions at the superconductor-antiferromagnet interface. Wave
functions for quasiparticles with energies below the
superconducting gap decay exponentially with increasing distance
from the interface. Solving the Andreev equations, one can easily
find the standard two-component solutions $\tilde \psi_{\sigma,
\pm}^{r} (\bm{k}_F)= \left({\tilde u}^r_{\sigma,\pm}(\bm k_F),
{\tilde v}^r_{\bar\sigma,\pm} (\bm k_F)\right)$ taken in the right
half-space of the boundary:
\begin{equation}
\tilde \psi_{\sigma,\pm}^{r}(\bm{k}_F)=C_\pm
{E\pm i\sqrt{\Delta_s^2-E^2}\choose -\Delta_s} \enspace .
\label{pm}
\end{equation}

Reflection amplitudes for electrons $r^e_\sigma$ and holes
$r^h_\sigma$, which enter the quasiclassical boundary conditions,
are taken for the normal-metal state of the superconducting region
at the Fermi surface. An AF/N boundary is impenetrable for
normal-metal quasiparticles with energies below the
antiferromagnetic gap, even in the absence of any interface
potentials (i.e. for a transparent interface). Hence, the
corresponding transmission amplitudes vanish and the complex
reflection amplitudes for electrons and holes have unit modulus:
$|r^e(\bm{k}_F)|=|r^h( \bm{k}_F)|=1$. Further, if $\tilde
\psi_{\sigma}^{r}(\bm{k}_F)$ represents an outgoing solution in
the case of a (110) interface, then
$\tilde\psi_{\sigma}^{r}(\bm{k}_F+ \bm{Q})$ is the incoming
solution, and vice versa. Here, $\bm{Q}=(\pm \sqrt{2\pi},0)$ only
has nonzero $x$ component. For this reason, it is convenient to
formulate the boundary conditions for the (110) interface
indicating explicitly only the $k_y$ component:
\begin{equation}
{\tilde\psi}^{r}_{\sigma,+}({k}_{y}) = \left (r^e_\sigma(k_y)
\frac{\displaystyle 1+ \hat\tau_z}{\displaystyle 2}+r^h_\sigma(k_y)
\frac{\displaystyle 1 - \hat\tau_z}{\displaystyle 2} \right)
\tilde{\psi}^{r}_{\sigma,-}({k}_{y}) \enspace.
\label{bc110}
\end{equation}

A relation between $r^e_\sigma(\bm{k}_F)$ and
$r^h_\sigma(\bm{k}_F)$ follows from the fact that the quantity
$-[{u}_{-\sigma}(\bm k_F)]^*$ represents the wave function
${v}_{\bar\sigma}(\bm k_F)$ for a hole with energy $-E$ from one
spin sub-band if ${u}_{\sigma}(\bm k_F)$ is the wave function for
an electron with energy $E$ from the other spin sub-band. Under
the condition $\Delta\ll t$, one can consider the reflection
amplitudes at subgap energies $\pm E$ as taken at the Fermi
surface. Accounting additionally for the fact that incoming and
outgoing waves for holes are interchanged as compared with the
case of electrons, we find for normal-metal electrons and holes at
the (110) AF/N interface: $r^{h}_{\sigma}=
\left(r^{e^*}_{\bar\sigma}\right)^{-1}$. This condition simplifies
because of the equalities $|r^e(\bm{k}_F)|=|r^h( \bm{k}_F)| =1$:
\, $r^{h}_{\sigma}=r^{e}_{-\sigma}$. Eventually, applying the
boundary conditions (\ref{bc110}) to the solution (\ref{pm}), we
get
\begin{equation}
\left\{\!\!\begin{array}{l}
C_+\!\left(\!E\!+i\sqrt{\Delta_s^2-\!E^2}\right)\!=r^e_\sigma(\bm{k}_F)
C_-\!\left(\!E\!-i\sqrt{\Delta_s^2-\!E^2}\right)\\
C_+=r^e_{-\sigma}(\bm{k}_F)C_- \enspace .
\end{array}
\right. \label{C}
\end{equation}
This equation determines the energies of the bound states at a
(110) AF/sSC interface. In order to present the energy in a
convenient form, we divide the reflection amplitude into two parts
$r^e_\sigma(\bm{k}_F)=r_{sp}+ r_{Q,\sigma}$, which are symmetric
and antisymmetric with respect to spin inversion, respectively.
The first part $r_{sp}$ is actually spin-independent and can be
considered as the contribution to the reflection amplitude from
the specular reflection channel. The contribution to the
reflection amplitude from the spin-dependent $Q$ reflection
differs in symmetry from the specular reflection part. This is the
part antisymmetric in $\sigma$, $r_{Q,\sigma}$, which is an
imaginary quantity. The phase of $r_{Q,\sigma}$ differs by $\pi$
for spin up and down quasiparticles \cite{bobkova05}. It follows
from the definition given and the condition
$|r^e_\sigma(\bm{k}_F)|=1$ that reflection coefficients
$R_{sp}=|r_{sp}|^2$ and $R_{Q}=|r_{Q,\sigma} |^2$ satisfy the
relation $R_{sp}+R_{Q}=1$. With these quantities, we obtain from
Eq.(\ref{C}) the following bound state energies at the (110)
AF/sSC interface:
\begin{equation}
E=\pm\Delta_s\sqrt{R_{sp}(\bm{k}_F)}\enspace .
\label{bsafssc110}
\end{equation}

If a $d$-wave superconductor occupies the right half space instead of
the $s$-wave one, the expression for $\tilde\psi_{\sigma, \pm}^{r}(
\bm{k}_F)$ in Eq.(\ref{pm}) is modified with the substitution
$\Delta_s\to\pm\Delta_d(\bm{k}_F)$. As a result, we obtain the spectrum
of the bound states at the (110) AF/dSC interface:
\begin{equation}
E=\pm\Delta_d(\bm{k}_F)\sqrt{R_{Q}(\bm{k}_F)}\enspace .
\label{bsafdsc110}
\end{equation}

The specific expressions for the normal-state reflection
amplitudes can be found for the (110) AF/N interface along the
standard way by solving the Schr\"odinger equations for electrons
in the left and the right half-spaces and constructing incoming
and outgoing solutions in the normal-metal region, as well as
exponentially decaying solution in the antiferromagnetic region
for electrons with energies below the antiferromagnetic gap. The
reflection amplitude is fixed after substituting the bulk
solutions to the equations, taken at lines where the nearest
neighbor hopping mixes the normal-metal and the antiferromagnetic
regions:
\begin{eqnarray}
r_{Q,\sigma}(\bm{k}_f)=-i\sigma\left[1+\left(\dfrac{ma}{\sqrt{2}
v_{F,x}(k_y)}\right)^2\right]^{-1/2},\nonumber\\
\shoveleft{r_{sp}(\bm{k}_F)=\left[1+\left(\dfrac{\sqrt{2}v_{F,x}(k_y)}
{ma}\right)^2\right]^{-1/2}_{\enspace .}}
\label{rqsp}
\end{eqnarray}
where $v_{F,x}(k_y)=2\sqrt{2}ta\cos (k_y/\sqrt{2})$.

Substituting Eq.(\ref{rqsp}) into Eqs. (\ref{bsafssc110}),
(\ref{bsafdsc110}), we come to the results obtained in Ref.
\onlinecite{bobkova05}. We note now that the $(110)$ interface
represents a special situation, when the normal state Fermi
velocity for the half-filled lattice has only a $v_{F,x}$
component, which is perpendicular to the surface. Hence, outgoing
and incoming normal-metal quasiparticles move only along or
opposite to the surface normal. The quasiparticle trajectory of
this kind is intrinsic to the specular reflection channel.
However, $Q$ reflection takes place in this particular case along
the same quasiparticle trajectory. For this reason, specular and
$Q$ reflections make coherent contributions to the total
reflection amplitude $r^e_\sigma(\bm{k}_F)=r_{sp}+r_{Q, \sigma}$.
For different interface orientations, for example, for the $(100)$
interface, specular and $Q$ reflection takes place along different
trajectories and do not interfere with each other.

In the absence of interface potentials specular reflection of
quasiparticles arises entirely due to a mismatch of Fermi
velocities in the AF and the sSC. Since normal metal states are
presumably identical in the left and right halfspaces under the
conditions $\Delta\ll m,t$ the mismatch in the model is controlled
by the parameter $m/t$. As is seen from Eq.(\ref{rqsp}),
$r_{sp}\to 1$ when the antiferromagnetic gap is large, $|m|\gg
4t\cos k_y$. At the same time, $Q$ reflection becomes dominant in
the opposite limit $|m|\ll 4t\cos k_y$, taking place at energies
below the antiferromagnetic gap $|E|<|m|$. The quasiparticle bound
state energy (\ref{bsafssc110}) at the AF/sSC interface is almost
zero in the limit $|E|\ll \Delta_s$, whereas the bound state
(\ref{bsafdsc110}) at the AF/dSC interface lies very close to the
edge of the superconducting gap.

\section{Quasiparticle states at SC/AF interfaces}

In this section, we study quasiparticle spectra and the
corresponding local density of states in the vicinity of AF/SC
interfaces based on our self-consistent approach outlined above.
We consider (100) and (110) interfaces between AF and either
$s$-wave or $d$-wave superconductors. The coupling constants are
site dependent with:
\begin{eqnarray}
U_i&=&U~~~~~~\!\mbox{for}~~~~i \leq 0,\\
V_i&=&V~~~~~~\!\mbox{for}~~~~i > 0,
\end{eqnarray}
and zero elsewhere. In the following we will typically be studying
finite systems of length $N=100-200$ along the $x$ axis. The $k_y$
sum is performed explicitly by using 400 points in the Brillouin
zone. The interface is always positioned at the bond in the middle
of the system and the potential $h_i$ is only non-zero on the two
sites immediately adjacent to the interface. We apply open
boundary conditions equivalent to an impenetrable wall at each end
of the system.

\subsection{$s$-wave superconductor - antiferromagnet (100) interface}

\begin{figure}[!tbh]
\includegraphics[width=8.5cm,height=4.0cm]{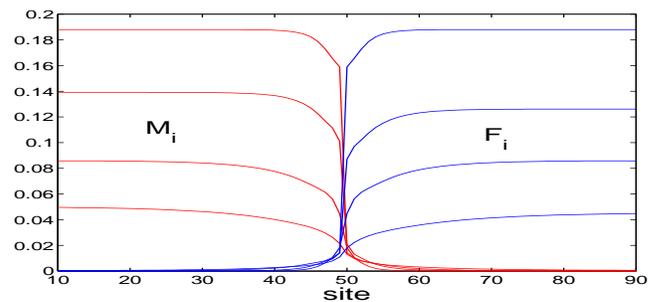}
\caption{(Color online) Self-consistent spatial dependence of the
absolute values of the magnetization $M_i$ and the pairing
amplitude $F_{ii}$ near a (100) interface. Parameters: $\mu=0$
and, from top to bottom, ($U=2.0t$,$V=2.0t$), ($U=1.6t$,$V=1.5t$),
($U=1.2t$,$V=1.2t$), and
($U=0.93t$,$V=0.9t$).\label{fieldsAFssc100}}
\end{figure}

We begin with the AF/sSC (100) situation since $Q$ reflection is
expected to lead to low-energy bound states for this particular
interface orientation\cite{bobkova05}. As is well known,
nonmagnetic interfaces do not break $s$-wave Cooper pairs in their
vicinity. In contrast, ferromagnetic as well as antiferromagnetic
interfaces are pair breaking, in general, even for $s$-wave
superconductors since they break time-reversal symmetry. This is
analogous to the difference between effects of nonmagnetic and
magnetic impurities in $s$-wave superconductors. In Fig.
\ref{fieldsAFssc100} we show the self-consistent suppression of
the magnetization $M_i$ and the pairing amplitude $F_{ii}$ near
the (100) interface for various values of superconducting and
antiferromagnetic coupling constants.
As seen, the healing length on each side of the interface
decreases with increasing amplitude of the order parameter in
agreement with the behavior of the respective magnetic and
superconducting coherence lengths $\sim \hbar v_{F,x}/|m|,\, \hbar
v_{F,x}/\Delta_s$. We find a correlation between the strength of
the suppression of order parameters and the energy of the Andreev
bound state arising near the AF/sSC interfaces. The lower the
(positive) energy of the subgap state, the stronger the
suppression of both order parameters at the interface, at least in
simple cases which have been studied.

\begin{figure}[!tbh]
\hspace{-3.0cm}(a) $~~~~~~~~~~~~~~~~~~~~~~~~~~~~~~~~~~$ (b)
\hspace{+3.0cm}
\includegraphics[width=8.5cm]{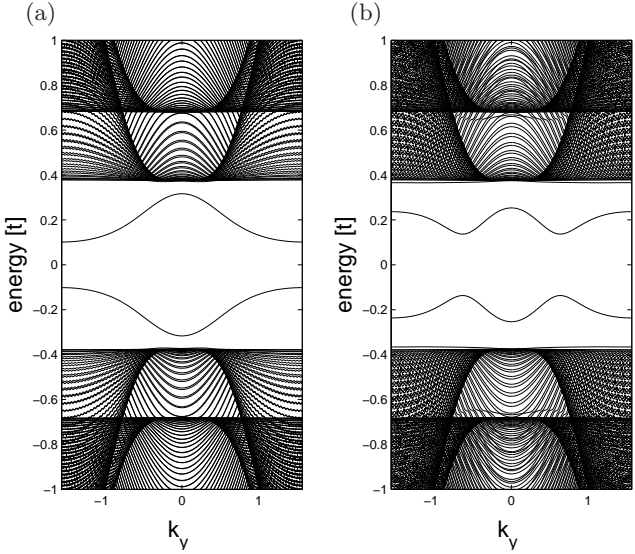}
\caption{Eigenvalues for the (100) AF/sSC interface as a function
of $k_y$ for $\mu=0.0$, $U=2.7t, V_s=2.0t$ and $h=0.0$ (a) and
$h=2.0t$ (b). Here, one sees explicitly the presence and
dispersion of the bound state band inside the
gap.\label{bandsAFssc100}}
\end{figure}

In Fig. \ref{bandsAFssc100} we plot the quasiparticle spectrum as
obtained from the eigenvalues of the BdG equations
(\ref{BdG1_ky_100})-(\ref{BdG2_ky_100}). Naturally, bound states
present at the interface show up inside the main gap of the
spectrum as a distinct band, which disperses with the momentum
component $k_y$ along the interface. The two gap edges seen in
Fig. \ref{bandsAFssc100} are associated with the superconducting
(lesser) and the antiferromagnetic (larger) gaps.

We have calculated the bound state spectrum also analytically,
assuming $\Delta_s\ll m,\, t$. Solving the Andreev equations for
the superconducting region and applying appropriate boundary
conditions, we obtain the two dispersive energies of quasiparticle
Andreev bound states for the (100) AF/sSC interface. The spectrum
is symmetric with respect to the zero level and can be described
with a reflection coefficient $R_{sp}(k_y)$ for quasiparticles in
the specular reflection channel: $E(k_y)=\pm
\Delta_s\sqrt{R_{sp}(k_y)}$. This expression actually coincides
with Eq.(\ref{bsafssc110}) derived in Sec. III for the $(110)$
interface although the explicit expression for the reflection
coefficient differ for the two orientations. A calculation of the
reflection coefficient $R_{sp}(k_y)$ for the (100) AF/sSC
interface in the absence of potential barriers leads to the
following explicit expression for energies of the Andreev bound
states:
\begin{eqnarray}\label{disp100sSC}
&&E(k_y)=\pm \Delta_s\times\nonumber\\
&& \left({\frac{A(k_y)+\sqrt{A^2(k_y)+4\left(\frac{\displaystyle
m}{ \displaystyle 2t}\right)^2}}{A(k_y)+2\sin^2 k_y +\sqrt{
A^2(k_y)+4\left(\frac{\displaystyle m}{\displaystyle 2t}
\right)^2}}}\right)^{1/2} \enspace,
\end{eqnarray}
where
\begin{equation} A(k_y)=\left(\frac{\displaystyle
m}{\displaystyle 2 t}\right)^2-\sin^2 k_y.
\end{equation}
The dispersion shown in Fig. \ref{bandsAFssc100}a can be verified
to agree well with the expression in Eq.(\ref{disp100sSC}) within
the accuracy $\pm(\Delta/t)^2$. Eq.(\ref{disp100sSC}) is very
similar to Eq.(7) of Ref. \onlinecite{bobkova205} and can be
obtained from there simply by substituting the magnetic $m$ and
$s$-wave $\Delta_s$ order parameters for the charge density wave
$W_s$ and $d$-wave $\Delta_d$ order parameters respectively.  As
follows from Eq.(\ref{disp100sSC}), the quasiparticle subgap state
becomes a dispersionless zero-energy state if one additionally
assumes $m\ll t$ and disregards terms less or the order of
$(m/t)^2$. This limiting case corresponds to the zero-energy
solution found from the quasiclassical Andreev equations applied
to both superconducting and antiferromagnetic regions under the
conditions $m,\,\Delta_s\ll t$ \cite{bobkova05}.

\begin{figure}[!tbh]
\hspace{-3.0cm}(a) $~~~~~~~~~~~~~~~~~~~~~~~~~~~~~~~~~~$ (b)
\hspace{+3.0cm}
\includegraphics[width=8.5cm]{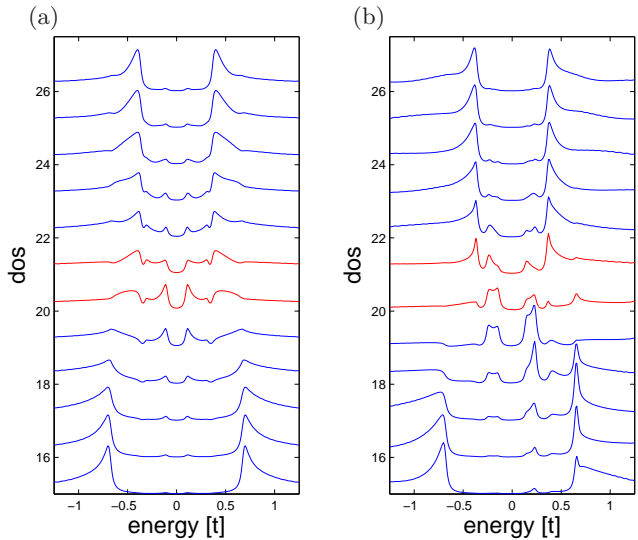}
\caption{(Color online) LDOS corresponding to the same parameters
as in Fig. \ref{bandsAFssc100}. The interface bound states result
in sub-gap peaks in the LDOS near the interface region. The two
center LDOS scans in both (a) and (b) are at the interface while
the top (bottom) five scans are into the SC (AF). The lines are
off-set for clarity.\label{ldosAFssc100}}
\end{figure}

The differential tunneling conductance measured, for instance, by
scanning tunneling microscopy (STM) experiments is proportional to
the local density of states\cite{tersoffhamann}. Therefore, it is
important to calculate the LDOS, given by
\begin{equation}\label{ldos}
N^{\alpha}_i(\omega)=-\frac{\mbox{Im}}{\pi}\sum_{nk_y\sigma} \left[
\frac{|u^{\alpha}_{n,i,\sigma}(k_y)|^2}{\omega-E_{nk_y\sigma}-i\Gamma} +
\frac{|v^{\alpha}_{n,i,\sigma}(k_y)|^2}{\omega+E_{nk_y\sigma}-i\Gamma} \right],
\end{equation}
where $\alpha=A,B$ indicates the magnetic sublattice and $\Gamma$
is an artificial broadening which in the following is set to
$\Gamma=0.02t$. For the (100) interface, we find
$N^{A}_i(\omega)=N^{B}_i(\omega)$. In plots of the resulting LDOS
we expect any bound states to result in additional peaks inside
the gap of the bulk AF and SC. These peaks should be localized
near the interface. This is indeed the case for the 100 AF/sSC
interface, as can be seen from Fig. \ref{ldosAFssc100}. Here, the
two center LDOS scans in both (a) and (b) are at the interface
while the top (bottom) five scans are moving into the SC (AF).

Additional potentials $h \neq 0$ present near the interface can
strongly enhance specular reflection at the expense of
Q-reflection. There are several consequences of the interface
potential for Andreev bound states present in the system. First,
potentials tend to suppress the bound states resulting from
Q-reflection and move their positions towards the gap edge. As
expected, in the limit $h \gg t$ we always find that the
Q-reflection bound states have been pushed into the continuum.
Second, in the regime where $h$ is of the order of $t$, we find
that the main effect of the specular reflection channel is to
cause a stronger dispersion of the bound state energy. One can
identify additional extrema in the wave vector dependence of the
bound state energy $E(k_y)$. A typical example is seen in Fig.
\ref{bandsAFssc100}b where $h=2.0t$. The new stationary points in
the dispersion lead to additional LDOS peaks near the interface as
seen in Fig. \ref{ldosAFssc100}b. In the LDOS we also see that the
particle-hole symmetry is broken when $h \neq 0$, whereas the
quasiparticle spectrum is still symmetric with respect to the
Fermi level. A similar asymmetry between positive and negative
bias in the LDOS will be present when starting from a
particle-hole asymmetric band, i.e. when $\mu \neq 0$. For the
sake of clarity, the results presented below are for the
particle-hole symmetric nested band where $\mu=0$ and any
asymmetry will only result from a non-zero interface potential
$h$.

\begin{figure}[!tbh]
\hspace{-3.0cm}(a) $~~~~~~~~~~~~~~~~~~~~~~~~~~~~~~~~~~$ (b)
\hspace{+3.0cm}
\includegraphics[width=8.5cm]{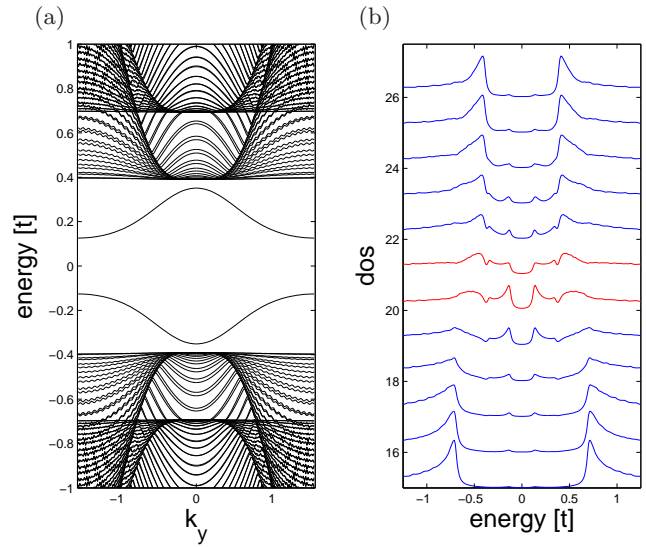}
\caption{(Color online) Non-self-consistent spectrum and LDOS
corresponding (roughly) to the same parameters as in Fig.
\ref{bandsAFssc100}a but with step-function spatial dependence of
$M_i$ and $F_{ii}$. These results are basically identical to those
shown in Fig. \ref{bandsAFssc100}a verifying that it is the
Q-reflection, not the order parameter suppression, that generates
the subgap states. \label{nsbandsandldos}}
\end{figure}

It is interesting to investigate the importance of the suppression
of the order parameters near the interface resulting from the
self-consistency. Fig. \ref{nsbandsandldos} shows the bands and
the corresponding LDOS for a non-self-consistent calculation with
step-function fields: $m^A_i=-m^B_i=m_0\Theta(-i)$ and
$\Delta_s=\Delta_0 \Theta(i)$ with $m_0=0.7t$ and $\Delta_0=0.4t$.
Clearly, the results are very similar to those shown for $h=0$ in
Fig. \ref{bandsAFssc100}a and Fig. \ref{ldosAFssc100}a,
respectively. This result applies also to the other interfaces
studied below: in general the self-consistency has only minor
effects on the bound states resulting from Q-reflection. It can,
on the other hand, depending on specific parameters, induce new
bound states close to the gap edge.

\begin{figure}[!tbh]
\includegraphics[width=8.5cm]{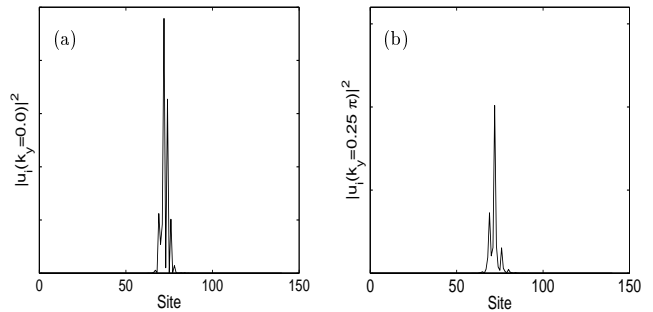}
\caption{Amplitude of $|u^{A}_i(k_y)|^2$ as a function of the
x-component of the unit cell $i$ corresponding to the bound state
in Fig. \ref{bandsAFssc100}a when $k_y=0.0$ (a) and $k_y=0.25\pi$
(b), respectively.\label{usquared}}
\end{figure}

As a final point in this section, we verify the bound nature of
the subgap states by showing explicitly the spatial dependence of
the eigenstates corresponding to the subgap band in e.g. Fig.
\ref{bandsAFssc100}a. This is done in Fig. \ref{usquared}, where
we plot $|u^{A}_i(k_y)|^2=|u^{B}_i(k_y)|^2$ as a function of the
x-component of the unit cell near the interface for $k_y=0.0$ and
$k_y=0.25 \pi$. Clearly, the wavefunctions are seen to be bound to
the interface region.

\subsection{$s$-wave superconductor - antiferromagnet (110) interface}

Unlike the (100) interface studied in the preceding section, in
the (110) case all sites at the interface belong to the same
sublattice, so that the interface layer (chain) itself is
ferromagnetically ordered. Up and down magnetizations alternate
only along the interface normal, i.e. the $x$-direction as seen
from Fig. \ref{110}. For this reason some characteristic
properties of AF/sSC (110) interfaces could naively be expected to
be reminiscent of the properties of superconductor-ferromagnetic
boundaries. For example, Cooper pair wave functions are known to
decay into the ferromagnet adjacent to the superconductor,
manifesting at the same time spatial oscillations
\cite{buz82,radovic}. The oscillations are known to be induced by
the difference between the momenta of spin up and down
quasiparticles with the same energy \cite{demler}. These
oscillations in the SC/F proximity effect are related to those in
the Fulde-Ferrell-Larkin-Ovchinnikov (FFLO) superconducting phase
\cite{fflo}.

\begin{figure}[!tbh]
\includegraphics[width=8.5cm,height=6.0cm]{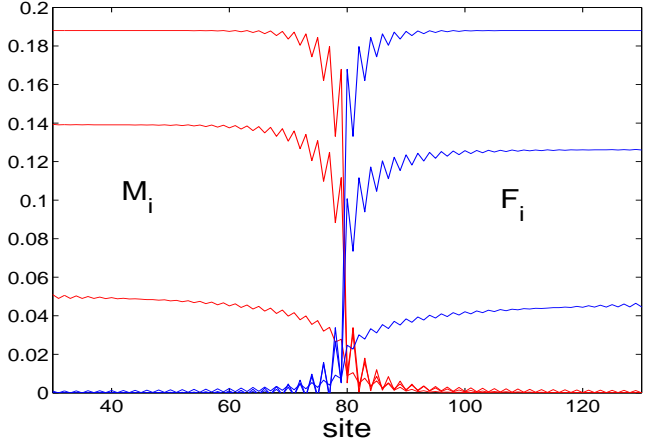}
\caption{(Color online) Spatial dependence of the self-consistent
results for the absolute value of the magnetization $M_i$ and
pairing amplitude $F_{ii}$ for the AF/sSC 110 interface
orientation. Parameters: $\mu=0$ and (top) $U=2.0t$, $V=2.0t$,
(middle) $U=1.6t$, $V=1.5t$, and (bottom) $U=0.93t$,
$V=0.9t$.\label{fieldsAFssc110}}
\end{figure}

In Fig. \ref{fieldsAFssc110} we show the obtained self-consistent
spatial profiles of the magnetization $M_i$ and the pairing
amplitude $F_{ii}$ near the (110) AF/sSC interface region for
various coupling strengths.
Both the magnetization and the pairing amplitude display an
oscillatory decaying behavior near the interface. The oscillations
have  even-odd character, i.e. they are not present within each
separate magnetic sublattice. Thus, the oscillations in Fig.
\ref{fieldsAFssc110} are not equivalent to  FFLO oscillations, but
simply induced by the AF staggered ordering. The characteristic
scales for suppression of the order parameters are seen to be the
corresponding coherence lengths of the antiferromagnet and the
superconductor. We find that an additional potential at the
interface decreases both the decay length and the amplitude of the
oscillations on both sides of the interface.

\begin{figure}[!tbh]
\hspace{-3.0cm}(a) $~~~~~~~~~~~~~~~~~~~~~~~~~~~~~~~~~~$ (b)
\hspace{+3.0cm}
\includegraphics[width=8.5cm]{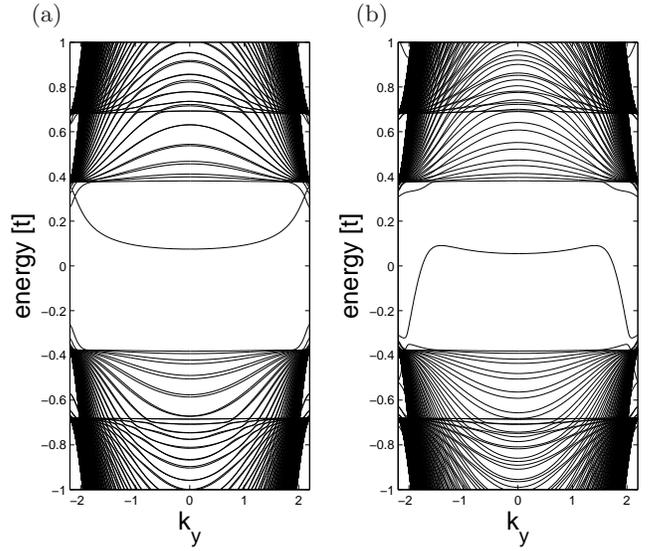}
\caption{Spin up eigenvalues resulting from the Bogoliubov-de
Gennes equations in the case of a 110 AF/sSC interface. The
parameters are $U=2.7t$, $V=2.0t$, and $h=0.0$ for
figure (a) and $h=1.0t$ for (b). The bands shown here correspond
to $(u_\uparrow, v_\downarrow)$. The bands associated with
$(u_\downarrow, v_\uparrow)$ are identical to the bands shown here
upon mirror reflection around $E=0$.\label{bandsAFssc110}}
\end{figure}

The dispersion of the subgap energies at (110) AF/sSC interfaces
are described analytically by Eqs.(\ref{bsafssc110}),
(\ref{rqsp}), within the non-self-consistent quasiclassical
approach. In Fig. \ref{bandsAFssc110}(a) we plot the numerically
determined self-consistent (spin up) eigenbands for the (110)
AF/sSC interface.
The dispersion of the main bound state can be checked to yield
excellent agreement with the result in Eq. (\ref{bsafssc110}).
Positive and negative energies correspond there to spin up and down
quasiparticles respectively. Thus, at low temperatures only spin
down quasiparticle Andreev states will be occupied. Strongly
spin-discriminated Andreev states together with the alternating
magnetization of chains, which are parallel to the (110)
interface, are at the origin of the even-odd oscillations of the
order parameters shown in Fig. \ref{fieldsAFssc110}. It is worth
noting that spin polarized Andreev states are compatible with
singlet spin structure of Cooper pairs \cite{bb02}. Whereas the
electron and the Andreev reflected hole belong to different spin
subbands and have opposite Zeeman energies, they possess identical
spin polarization. Since they have also almost opposite
velocities, they do not carry together any spin current except for
special cases \cite{bb04}.

At the edge of the Brillouin zone we find new high-energy subgap
states. By comparing to non-self-consistent calculations with
step-like spatial order parameter dependence, we have found that
these states are related to the self-consistent order parameter
suppression near the interface.  When including a potential at the
interface, the structure of the dispersion of the subgap states
becomes more complicated. In Fig. \ref{bandsAFssc110}b we show an
example where $h=t$. When the potential becomes too large, $h \gg
m,\Delta$, the specular reflection channel completely dominates
the Q-reflection and all bound states are pushed out of the gap
into the continuum.

\begin{figure}[!tbh]
\hspace{-3.0cm}(a) $~~~~~~~~~~~~~~~~~~~~~~~~~~~~~~~~~~$ (b)
\hspace{+3.0cm}
\includegraphics[width=8.5cm]{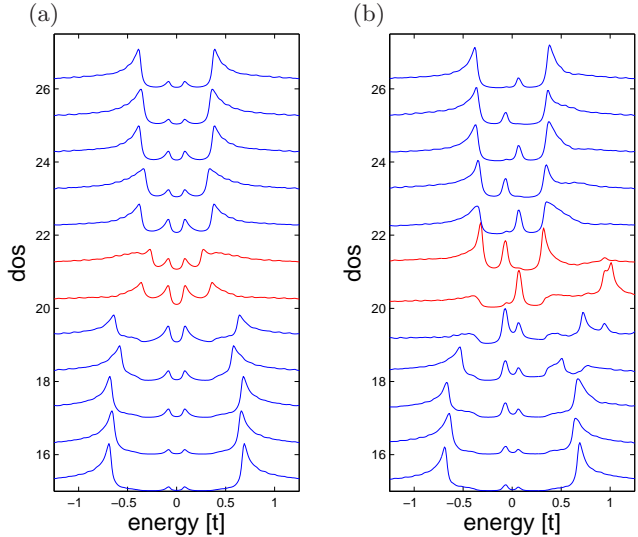}
\caption{(Color online) LDOS near the (110) AF/sSC interface
corresponding to the parameters used in Fig. \ref{bandsAFssc110}.
The two center LDOS scans in both (a) and (b) are at the interface
while the top (bottom) five scans are the LDOS upon moving into
the SC (AF). The lines are off-set for
clarity.\label{ldosAFssc110}}
\end{figure}

The spin-summed LDOS corresponding to the results shown in Fig.
\ref{bandsAFssc110}a and Fig. \ref{bandsAFssc110}b is shown in
Fig. \ref{ldosAFssc110}. Since in the (110) case sites $A$ and $B$
from the same cell are at different distances from the interface,
$N^{A}_i(\omega)$ and $N^{B}_i(\omega)$ differ from each other.
Each of Figs. \ref{bandsAFssc110}a, \ref{bandsAFssc110}b display
$N^{A}_i(\omega)$ and $N^{B}_i(\omega)$ jointly, i.e. shows the
LDOS for all sites along the x-axis. As expected the bound states
again show up as peaks in the LDOS close to the interface. The
states at the Brillouin zone edge can give rise to higher energy
peaks. The main influence of the potential scattering is to break
the particle-hole symmetry present in the LDOS in Fig.
\ref{ldosAFssc110}a. In particular, as seen from Fig.
\ref{ldosAFssc110}b, the presence of a potential $h$ can lead to
distinct even/odd amplitude modulations of the bound state LDOS
peaks into the superconductor.

\subsection{$d$-wave superconductor - antiferromagnet (100) interface}

The electronic structure of interfaces formed by $d$-wave
superconductors and antiferromagnets is an important problem
relevant to e.g. the high temperature superconductors where AF and
dSC order dominate the phase diagram. A qualitative difference
between the properties of AF/sSC and AF/dSC interfaces arises from
the fact that the quasiclassical $d$-wave order parameter
$\Delta_d^j(\bm{k}_{f})$ changes sign in a $Q$ reflection event
for any interface orientation:
$\Delta_d(\bm{k}_F+\bm{Q})=-\Delta_d(\bm{k}_F)$. This change of
sign can strongly weaken the effect of phase difference of
reflection coefficients for spin up and down quasiparticles, which
is close to $\pi$ in the limit $\Delta, \,m\ll t$.  For this
reason, low-energy interface bound states $E_B(\bm{k}_F)\ll {\rm
min}\{m,\Delta_{\rm bulk}(\bm{k}_F)\}$ existing at AF/sSC
interfaces under the conditions $\Delta_s,\, m\ll t $, do not
exist at AF/dSC interfaces with arbitrary orientation
\cite{bobkova05}. In the present section we demonstrate results of
self-consistent numerical calculations for the spatial profiles of
superconducting and antiferromagnetic order parameters, the
quasiparticle spectrum, and the associated LDOS in the vicinity of
$(100)$ AF/dSC interfaces.

\begin{figure}[!tbh]
\includegraphics[width=8.5cm,height=4.0cm]{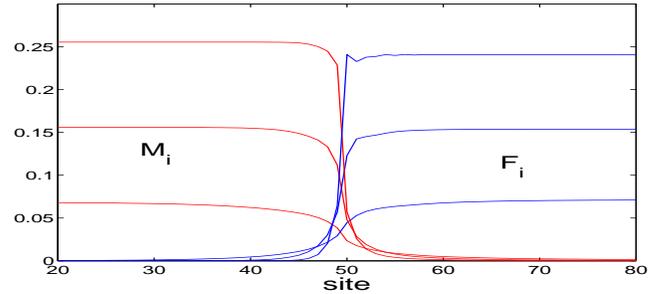}
\caption{(Color online) Spatial dependence of the self-consistent
results for the absolute value of the magnetization $M_i$ and
pairing amplitude $F_i$ for the AF/dSC 100 orientation.
Parameters: $\mu=0$, and (top to bottom) $U=2.7t$, $V=2.0t$;
$U=1.7t$, $V=1.0t$; and $U=1.1t$, $V=0.5t$.\label{fieldsAFdsc100}}
\end{figure}

Fig. \ref{fieldsAFdsc100} shows a typical result for the
suppression of $M_i$ and $F_i$ near the interface. Here, we plot
the $d$-wave pairing amplitude defined on-site in the usual way
from the four surrounding links $F_i=\frac{1}{4} \left( F_{i,i+a}
+ F_{i,i-a} - F_{i,i+b} - F_{i-b} \right)$.
Comparing Fig. \ref{fieldsAFdsc100} and Fig. \ref{fieldsAFssc100},
one can notice weaker pair breaking effects on the scale of the
respective coherence length for the AF/dSC (100) interface
compared to the AF/sSC interface. This is associated with lower
bound state energies at AF/sSC interfaces. We find similar results
for the (110) AF/dSC interface studied in the next section.
Contrary to the (110) AF/sSC case, the order parameters exhibit a
rather smooth suppression near a (100) or (110) AF/dSC interface,
i.e. the order parameter oscillations are absent. This correlates
with the absence of a spin discrimination in the quasiparticle
subgap spectrum generated by the (110) AF/dSC boundary.

\begin{figure}[!tbh]
\hspace{-3.0cm}(a) $~~~~~~~~~~~~~~~~~~~~~~~~~~~~~~~~~~$ (b)
\hspace{+3.0cm}
\includegraphics[width=8.5cm]{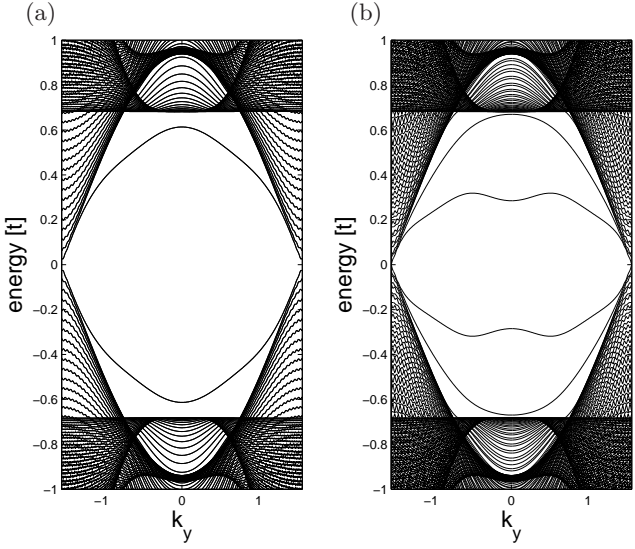}
\caption{Eigenbands for the AF/dSC 100 interface as a function of
$k_y$. Parameters: $U=2.7t$, $V=2.0t$, $\mu=0$ and $h=0.0$ (a) and
$h=1.0t$ (right).\label{bands1AFdsc100}}
\end{figure}

The eigenbands for two different parameter sets are shown in Fig.
\ref{bands1AFdsc100}. The left figure shows the result without any
potential at the interface. Here, a subgap band exists close to
the edge of the continuum. We find that the larger the ratio
$m/\Delta$, the closer the subgap band is to the gap edge. This
agrees with Ref. \onlinecite{bobkova05}, where it was demonstrated
that there are no interface bound states on AF/dSC (100)
interfaces in the limit $m\gg\Delta$. Only upon decreasing the
ratio $m/\Delta$ does a bound state gets peeled off the continuum,
as seen in Fig. \ref{bands1AFdsc100}a. The band is degenerate and
will be split by a non-zero potential $h \neq 0$ as seen in Fig.
\ref{bands1AFdsc100}b. This is accompanied by the appearance of
additional extrema in the dispersion of a lower band. Although
with increasing $h$ a lower band first becomes closer to the Fermi
level than for $h=0$, its position depends nonmonotonically on $h$
and never approaches  zero energy. When $h \gg t$ both bound state
bands are pushed into the continuum. The results of
non-self-consistent calculations are very similar to the results
shown in Fig. \ref{bands1AFdsc100}: it is the Q-reflection channel
that induces the bound states at the AF/dSC interface, not the
order parameter suppression.

\begin{figure}[!tbh]
\hspace{-3.0cm}(a) $~~~~~~~~~~~~~~~~~~~~~~~~~~~~~~~~~~$ (b)
\hspace{+3.0cm}
\includegraphics[width=8.5cm]{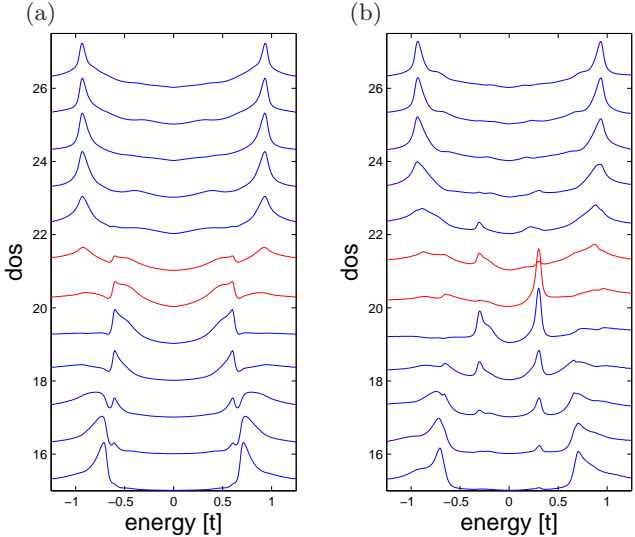}
\caption{(Color online) LDOS corresponding to the plots shown in
Fig. \ref{bands1AFdsc100}. The two center LDOS scans in both (a)
and (b) are at the interface while the top (bottom) five scans
shows the LDOS when moving into the SC (AF).\label{ldosAFdsc100}}
\end{figure}

The LDOS corresponding the same parameters used in Fig.
\ref{bands1AFdsc100} is shown in Fig. \ref{ldosAFdsc100}.
Again the new LDOS peaks arising from the bound states are
sensitive to the presence of the interface potential $h$. When
$h=0$ the bands in Fig. \ref{bands1AFdsc100}a close to the
continuum edge gives rise to LDOS peaks near the coherence peaks.
Experimentally, it may be a challenge to distinguish these bound
state peaks from the coherence peaks of the bulk dSC. However, for
the case when $h=1.0t$, we see that the lower band from Fig.
\ref{bands1AFdsc100}b results in sharp LDOS peaks near
the interface in the intermediate region of subgap energies.

\subsection{$d$-wave superconductor - antiferromagnet (110) interface}

As is well-known, a zero-energy Andreev bound state exists at the
(110) insulator I/dSC interface generated by the sign-reversal of
the gap function as seen by a quasi-particle specularly scattered
off the surface\cite{zesref1,zesref2,zesref3}. This state has been
observed in the differential tunneling conductance as a conductance
peak at zero bias \cite{zesexpref1,zesexpref2,cov97,alf97,ek97,ap98,%
alf981,alf982,sin98,wei98,apr99,deutch99,cov00,zesexpref3,greene03}.
In this section we study the AF/dSC interface with (110) orientation.

\begin{figure}[!tbh]
\hspace{-3.0cm}(a) $~~~~~~~~~~~~~~~~~~~~~~~~~~~~~~~~~~$ (b)
\hspace{+3.0cm}
\includegraphics[width=8.5cm]{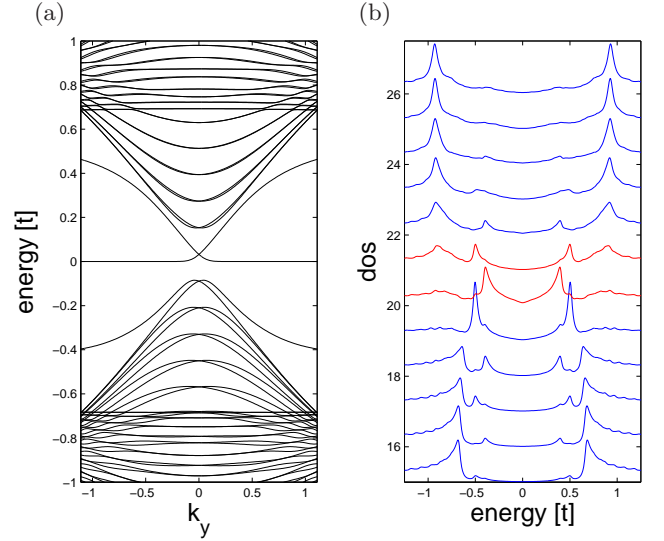}
\caption{(Color online) Bands and LDOS for the 110 AF/dSC
interface with $U=2.7t$, $V=2.0t$ and $h=0.0$. The low-energy
bands are dominated by the gap node at $k_y=0$, the Q-reflection
bounds states and a zero energy state. The latter state is located
at the dSC/I boundary at the right-most the end of our system
where the open boundary conditions operate as a hard
wall.\label{bandsandldosAFdsc110}}
\end{figure}

In Fig. \ref{bandsandldosAFdsc110} we plot the (spin up) bands and
the corresponding LDOS when $U=2.7t$, $V=2.0t$ and $h=0.0$.
Since the open boundary conditions at the edges of our finite
system are equivalent to a hard wall, a zero energy state (ZES) is
present at the superconducting end of our system. This state,
clearly seen in Fig. \ref{bandsandldosAFdsc110}a, is not
associated with the AF/dSC interface in which we are interested.
It mixes with the bound states resulting from the Q-reflection at
the interface, however, and causes small deviations from the
expected mirror symmetry (through $E=0$) of the bound state bands.
In the limit of an infinite system, this spurious effect
disappears and the continuum gap node closes at $k_y=0$.

The dispersion of the subgap states in the absence of potential
barriers is described analytically in Eqs.(\ref{bsafdsc110}),
(\ref{rqsp}) assuming spatially constant order parameters
$\Delta_d(\bm {k}_F)\ll m,\, t$. These states should move to lower
energy as $m$ increases. For our band $v_{F,x}=2\sqrt{2}ta
\cos(k_y/\sqrt{2})$ and again it can be easily verified that the
functional form of the self-consistent subgap band in Fig.
\ref{bandsandldosAFdsc110}a agrees well with Eqs.
(\ref{bsafdsc110}), (\ref{rqsp}).

The presence of a ZES at (110) I/dSC interfaces, makes it
interesting to plot the LDOS as a function of increased interface
potential $h$.
\begin{figure}[!tbh]
\hspace{-3.0cm}(a) $~~~~~~~~~~~~~~~~~~~~~~~~~~~~~~~~~~$ (b)
\hspace{+3.0cm}
\includegraphics[width=8.5cm]{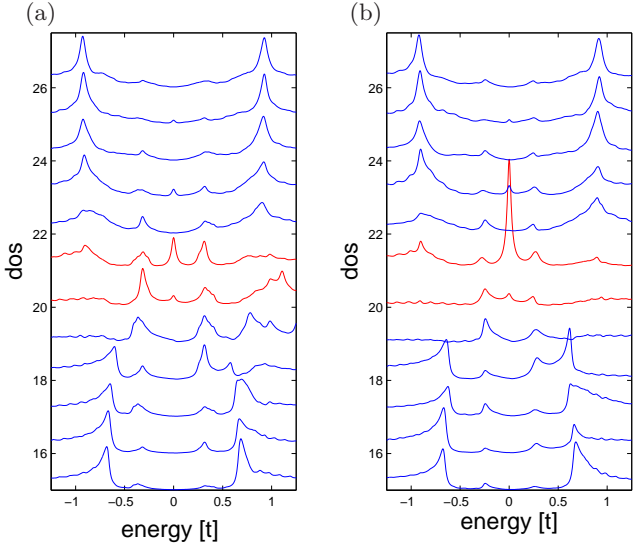}
\caption{(Color online) LDOS for the 110 AF/dSC interface with
$\mu=0$, $U=2.7t$, $V=2.0t$ and $h=1.0t$ (a) and $h=2.0t$ (b).
Again, the two center LDOS scans in both (a) and (b) are at the
interface while the top (bottom) five scans shows the LDOS when
moving into the SC (AF). Here, one clearly sees the continuous
evolution of the well-known zero energy state with increasing
interface potential $h$.\label{ldosAFdsc110}}
\end{figure}
Fig. \ref{ldosAFdsc110} shows the evolution of the LDOS when
approaching the interface in two cases where $h=t$ (a) and $h=2t$
(b). The details of the dispersion of the subgap states are
sensitive to the strength of the potential barrier at the
interface, and the resulting LDOS will strongly depend on $h$. In
the limit where $h \gg t$ the low-energy LDOS near the interface
will be dominated by the ZES. However, in the regime where $h \sim
t$, the ZES coexists with the Q-reflection bound states as is
evident from Fig. \ref{ldosAFdsc110}. This finding is relevant for
the discussion of possible surface induced magnetization near
I/dSC interfaces in cuprate superconductors
\cite{ohashi99,honerkamp,andersen02}. In the case of finite
$M_i$ near a (110) dSC surface, we would expect small side-band
peaks originating from the Q-reflection as seen in Fig.
\ref{ldosAFdsc110}. To the best of our knowledge no such features
have yet been identified in  experiments.

\subsection{Transfer matrix method}

The existence and dispersion of bound states at interfaces between
SC and AF can also be conveniently formulated within a transfer
matrix method designed to locate the bound states from their
defining property: spatially decaying wavefunctions. Below, we use
the same transfer matrix formalism presented in Ref.
\onlinecite{andersen02}. In this method, one introduces a
($q,\epsilon$)-dependent matrix $T\left(i+1,i\right)$ defined by
\begin{equation}
\Psi\left(i+1\right) = T\left(i+1,i\right) \Psi\left(i\right),
\end{equation}
which transfers the spinor $\Psi$ from site $i$ to site $i+1$. For
a model with nearest neighbor coupling $\Psi$ takes the explicit
form $\Psi \left(i\right) = \left( \psi \left(i\right),\psi
\left(i-1\right) \right)$ where
\begin{equation}
\psi \left(i\right) = \left( u_{k_y\sigma}\left(i\right),
  v_{k_y\sigma}\left(i\right), u_{k_y+\pi\sigma}\left(i\right),
  v_{k_y+\pi\sigma}\left(i\right) \right).
\end{equation}
The associated $8\times8$ transfer matrix has the general form
\begin{equation}
T\left(i+1,i\right) = \left(
\begin{array}{cc}
A & B \\
1 & 0
\end{array}
\right)
\end{equation}
where $A$ ($B$) denotes the $4\times4$ coefficient-matrix
connecting $\psi\left(i+1\right)$ and $\psi\left(i\right)$
($\psi\left(i-1\right)$) determined from the Bogoliubov-de Gennes
equations (\ref{BdG}).

For a typical interface there will be three distinct transfer
matrices; one in the bulk magnetic region $T_M$, one in the bulk
SC region $T_{SC}$ and one associated with transfer through the
interface $T_I$. By diagonalizing $T_M$ and $T_{SC}$ one
determines whether eigenstates decay, grow or propagate from the
interface depending on whether the eigenvalues are less than,
larger or equal to one, respectively. Here, decaying and growing
refer to propagation along the $x$-axis.

Let $PE_M$ denote the matrix obtained after propagating the
eigenvectors of the bulk magnetic transfer matrix through the
interface. Then we can define a matrix $A$ given by
\begin{equation}
PE_M = E_{SC} \cdot A
\end{equation}
where $E_{SC}$ is the matrix containing the eigenvectors of the
bulk superconducting region as column vectors and dot indicates
matrix multiplication. Let $S_g^{M}$ and $S_g^{SC}$ denote the
subspace of growing eigenstates of $PE_M$ and $E_{SC}$
respectively. Consider the following linear combination of the
{\sl growing} states of $PE_M$
\begin{eqnarray}\label{olebole}
\sum_{i\in S_g^{m}} \beta_i |PE_{M}i> &=& \sum_{i\in S_g^{m}}
\sum_{j
  \in S_g^{SC}} \beta_i A_{ji} |E_{SC}j> \\ \nonumber
&=& \sum_{j\in S_g^{SC}} \left( \sum_{i\in S_g^{M}} A_{ji} \beta_i
\right) |E_{SC}j>
\end{eqnarray}
>From equation (\ref{olebole}), it is evident that to have a bound
state at the interface, the vector $\beta$ must belong to the null
space of the reduced matrix $A_r$, defined to be the $S_g^{SC}
\times S_g^{M}$ upper left part of the original matrix $A$. This
follows since the matrices $PE_M$ and $E_{SC}$ were chosen
to have the eigenstates with the largest eigenvalues as
column vectors to the left. Thus, when the two subspaces
$S_g^{SC}$ and $S_g^M$ have the same dimension, a bound state at
the interface is characterized by the vanishing of the determinant
of $A_r$
\begin{equation}\label{det}
\mbox{Bound state criterion:}~~~~~\det A_r = 0.
\end{equation}
This is the criterion used previously by Andersen and Hedegard to
study the splitting of the ZES near dSC/AF interfaces
\cite{andersen02}.
\begin{figure}[!tbh]
\hspace{-3.0cm}(a) $~~~~~~~~~~~~~~~~~~~~~~~~~~~~~~~~~~$ (b)
\hspace{+3.0cm}
\includegraphics[width=8.5cm,height=4.0cm]{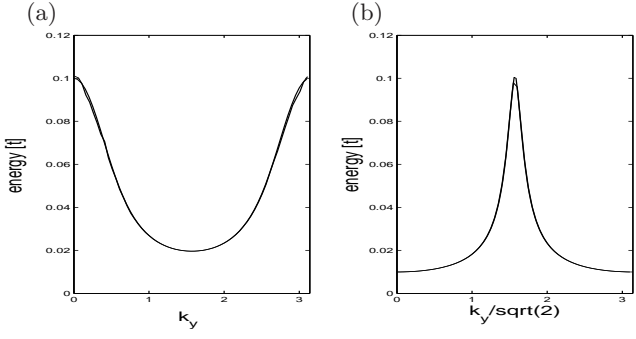}
\caption{Comparison of the dispersion obtained for the 100 (a) and
110 (b) AF/sSC interface form the transfer matrix method and the
result of Eq. (\ref{disp100sSC}) (a) or Eqs. (\ref{bsafssc110}),
(\ref{rqsp}) (b), respectively. There appears to be only one plot in
each figure here because of the excellent agreement between these two
methods. \label{dispcomp}}
\end{figure}

In Fig. \ref{dispcomp}, we show the results of Eq. (\ref{det}) for
the 100 (a) and 110 (b) AF/sSC interface, respectively. Here, we
used a nested band $\mu=0.0$ with $M=0.4$ and $\Delta=0.1t$. In
both graphs we also plot the curves given by the analytical
results in Eq. (\ref{disp100sSC}) and Eqs. (\ref{bsafssc110}),
(\ref{rqsp}) respectively. The (almost) complete overlap of the
curves reveal that the transfer matrix method captures the
low-energy bound states, and that these have the same dispersion
as discussed in the previous sections. Small deviations are seen
near the Brillouin zone edges. This is expected since the Fermi
velocity $v_F$ vanishes there and the quasiclassical approximation
used to derive Eqs. (\ref{disp100sSC}) and (\ref{bsafssc110})
becomes less reliable.

For the AF/dSC interfaces we find similar agreement between the
analytical results and the transfer matrix method.

\section{The AF/N/AF junction}

Discrete quasiparticle bound states below the AF gap induced by Q
reflection processes at AF/N interfaces, exist in a planar AF/N/AF
junction analogous to Andreev bound states in SC/N/SC systems
\cite{bobkova05}. Here, we present the analytical solution of the
problem for the simple case $\Delta_s, m\ll v_{F,x}/a$, when the
sublattice quasiclassical approximation applies well to both the
superconductor and the antiferromagnet. Analogous to the Andreev
equations (\ref{AndrkQ}), we formulate Schr\"odinger equations for
electrons with the quasiclassical approximation in the $(\bm{k}_F,
\bm{k}_F+\bm{Q})$ representation:
\begin{multline}
\left(-iv_{F,x}(\bm{k}_{F})\dfrac{\partial}{\partial x}-\mu-E\right)
\tilde{u}_{\sigma,\bm{k}_{F}}(x)
\\ \shoveright{+\sigma m_\pm \tilde{u}_{\sigma,\bm{k}_{F}+\bm{Q}}(x)=0
\, ,} \\
\shoveleft{\left(iv_{F,x}(\bm{k}_{F})\dfrac{\partial}{\partial x}
-\mu-E\right)\tilde{u}_{\sigma, \bm{k}_{F}+\bm{Q}}(x)}\\
+\sigma m_\pm \tilde{u}_{\sigma,\bm{k}_{F}}(x)=0 \, .
\label{schr}
\end{multline}
The relation $v_{F,x}(\bm{k}_{F}+\bm{Q})=-v_{F,x}(\bm{k}_{F})$
is taken into account in Eq.(\ref{schr}).

Let $\theta$ be a misorientation angle between two magnetic order
parameters $\bm{m}_\pm$ in the right $x>d/2$ and the left $x<-d/2$
antiferromagnetic halfspaces. Consider the magnetizations lying in
the $yz$ plane, which is perpendicular to the interface normal.
The quasiparticle spin will be not a good quantum number in the
case $\theta\ne 0, \pi$, so that we have to explicitly introduce
spin coordinates. We choose the global quantization axis along the
magnetization in the left antiferromagnet. Since Eqs.(\ref{schr})
are written for the quantization axis taken along the order
parameters $m_{\pm}>0$, the solutions in the right antiferromagnet
should be rotated by the angle $\theta$ around the $x$ axis in
spin space before matching them with the corresponding solution in
the normal metal region at $x=d/2$. After the rotation, the
low-energy solutions of Eqs.(\ref{schr}) in the right
antiferromagnetic halfspace can be written as
\begin{widetext}
\begin{equation}
\left[D_{+,\uparrow} \left(\begin{array}{c}
\cos(\theta/2)e^{\displaystyle i{\rm sgn}(v_{F,x})\varphi_+}\\ \\
i\sin(\theta/2)e^{\displaystyle i{\rm sgn}(v_{F,x})\varphi_+}\\ \\
\cos(\theta/2)\\
i\sin(\theta/2)\\
\end{array}\right)
+
D_{+,\downarrow} \left(\begin{array}{c}
-i\sin(\theta/2)e^{\displaystyle i{\rm sgn}(v_{F,x})\varphi_+}\\ \\
-\cos(\theta/2)e^{\displaystyle i{\rm sgn}(v_{F,x})\varphi_+}\\ \\
i\sin(\theta/2)\\
\cos(\theta/2)\\
\end{array}\right)\right]
\exp\left(-\dfrac{\sqrt{m_+^2-(E+\mu)^2}}{|v_{F,x}|}
x\right)\, .
\label{pmd2}
\end{equation}
\end{widetext}
Here, we have introduced the notation
\begin{equation}
e^{\displaystyle i\varphi_{\pm}(E)}=\dfrac{1}{m_\pm}\left(E+\mu
+i\sqrt{m^2_\pm-(E+\mu)^2}\right) \, .
\label{varphipm}
\end{equation}
The upper two lines in Eq.(\ref{pmd2}) describe spin up and down
amplitudes with the momentum $\bm{k}_F$ and the lower two lines
show the relative values of spin up and down amplitudes with the
momentum $\bm{k}_F+\bm{Q}$. The solutions for the left
antiferromagnetic halfspace can be obtained from Eq.(\ref{pmd2})
after the substitutions $\theta\to0$, $m_+\to m_-$,
$\varphi_+\to-\varphi_-$, $D_{+,\uparrow (\downarrow)}\to
D_{-,\uparrow (\downarrow)}$, $x\to -x$.

In the normal-metal region $-d/2<x<d/2$ the magnetization vanishes
and the solutions of Eqs.(\ref{schr}) take the form
\begin{widetext}
\begin{equation}
\left(\begin{array}{c} \tilde u_{\uparrow,\bm{k}_{F}}(x) \\
\tilde u_{\downarrow,\bm{k}_{F}}(x)\\ \tilde u_{\uparrow,\bm{k}_{F}
+\bm{Q}}(x)\\ \tilde u_{\downarrow,\bm{k}_{F}+\bm{Q}}(x)\end{array}\right)
=\left[C_1\left(\begin{array}{c} 1\\ 0\\0\\0\end{array}\right)+
C_3\left(\begin{array}{c} 0\\ 1\\0\\0\end{array}\right)\right]
e^{\displaystyle ix(E+\mu)/v_{F,x}}+ \left[C_2\left(\begin{array}{c}
0\\ 1\\0\\0\end{array}\right)+C_4\left(\begin{array}{c} 0\\0\\0\\1\end{array}
\right)\right]e^{\displaystyle -ix(E+\mu)/v_{F,x}} \, .
\label{pmmpd2}
\end{equation}
\end{widetext}
In the absence of potential barriers, the quasiclassical solutions
are continuous across the interfaces. Matching the solutions
(\ref{pmd2}), (\ref{pmmpd2}) at $x=\pm d/2$, we obtain the
following equation for bound state energies
\begin{equation}
E^\pm_n=\dfrac{|v_{F,x}|}{2d}\biggl(\varphi_{+}(E^\pm_n)+
\varphi_{-}(E^\pm_n)\pm\theta+2\pi n\biggr)-\mu \enspace ,
\label{En}
\end{equation}
where $n=0,\pm1,\pm2,\dots$.

The dependence of the bound state energies on the particular
values $m_\pm$ disappears for low-energy states in the almost
half-filled lattice. Indeed, under the condition $|E_n+\mu|\ll
m_\pm$ one can take $\varphi_{\pm}\approx \pi/2$ and find from
Eq.(\ref{En}) the following low-energy equidistant spectrum
\begin{equation}
E^\pm_{n}(\bm{k}_F,\theta)=\dfrac{|v_{F,x}(\bm{k}_F)|}{2d}\left[2\pi\left(n+
\dfrac{1}{2}\right)\pm\theta\right]-\mu \, .
\label{equi}
\end{equation}
As seen from Eq.(\ref{equi}), $E_{n+1}-E_n=\pi |v_{F,x}|/d$ and
$E^\pm_n(\pi)-E_n(0)=\pm\pi |v_{F,x}|/2d$. The condition
$|E_n+\mu| \ll m_\pm$ is valid for sufficiently large thickness of
the normal metal region $d\gg(\hbar v_{F,x}/m)\sim\xi_m$, when
there can be many levels described by Eq.(\ref{equi}). The
zero-energy bound state appears, in particular, in the half-filled
lattice for $\theta=\pi$ when the relative phase of the
antiferromagnetic ordering differs by $\pi$ in the left and the
right halfspaces.

The spectrum (\ref{equi}) qualitatively differs from that in a
conventional 'particle in a box', i.e. a system of almost free
quasiparticles confined by impenetrable walls constituting a I/N/I
junction. The reason for the difference is associated with strong
correlations between electrons with momenta $\bm{k}_F$ and
$\bm{k}_F+\bm{Q}$, induced in the normal metal region by the
antiferromagnets, where equations for electrons with $\bm{k}_F$
and $\bm{k}_F+\bm{Q}$ are coupled (see Eqs. (\ref{schr})).

After the substitution $m_\pm\to\Delta$ the bound state energies
(\ref{En}) coincide with those obtained many years ago for SC/N/SC
systems (at $\mu=0$) with the phase difference $\theta$
\cite{kulik69}. The reason for this is seen at $\theta=0$ where
the quasiclassical equations Eqs.(\ref{schr}) for electrons with
momenta $\bm{k}_F$ and $\bm{k}_F+\bm{Q}$ coincide with the Andreev
equations with $\mu=0$ and real $\Delta$ after the substitutions
$\sigma m\to \Delta$,\, $\tilde u_{\sigma,\bm{k}_{F}
+\bm{Q}}(x)\to\tilde v_{\bar\sigma,\bm{k}_{F}}(x)$. It is
important for this property that electrons with the momentum
$\bm{k}_F+ \bm{Q}$ possess a reverse velocity (similar to holes
with the momentum $\bm{k}_F$) compared with electrons with the
momentum $\bm{k}_F$. In the presence of a misorientation angle,
one can transform Eqs.(\ref{schr}) into the corresponding Andreev
equations at $\mu=0$ with complex $\Delta$, if one uses a gauge
transformation after rotating all the quantities in Eq.
(\ref{schr}) over the angle $\theta$ in spin space.

In the following, we study the AF/N/AF junction from the solution
of the BdG equations (\ref{BdG}). For the coupling constants we
have the following simple spatial dependence
\begin{eqnarray}
U_i&=&U~~~~~~~~\mbox{for}~~~~|i| > \frac{d}{2},\\
V_i&=&0~~~~~~~~~\mbox{for all}~~i,
\end{eqnarray}
and $U=0$ within the normal region $|i| \leq \frac{d}{2}$.

\begin{figure}[!tbh]
\hspace{-3.0cm}(a) $~~~~~~~~~~~~~~~~~~~~~~~~~~~~~~~~~~$ (b)
\hspace{+3.0cm}
\includegraphics[width=8.5cm]{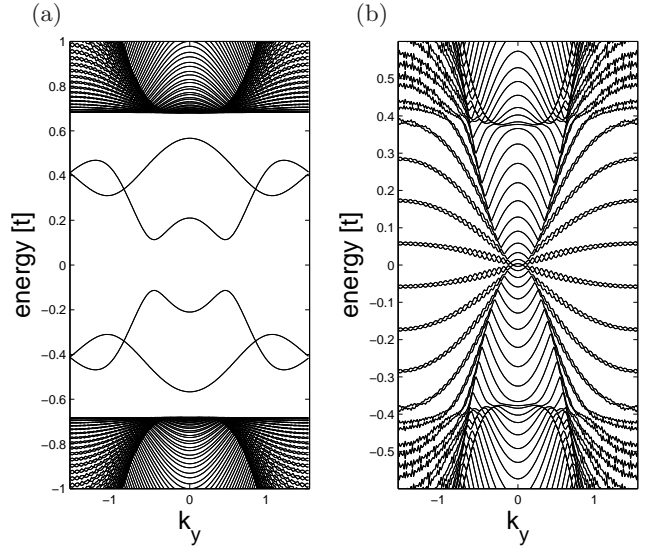}
\caption{Bound state eigenbands for a 100 AF/N/AF junction where
the length $d$ of the normal region N is either $d=4$ (a) or
$d=48$ (b) lattice sites. Parameters: $\mu=0$, $U=2.7t$ (a) and
$U=2.0t$ (b).\label{bandsAFNAF}}
\end{figure}

\begin{figure}[!tbh]
\hspace{-3.0cm}(a) $~~~~~~~~~~~~~~~~~~~~~~~~~~~~~~~~~~$ (b)
\hspace{+3.0cm}
\includegraphics[width=8.5cm]{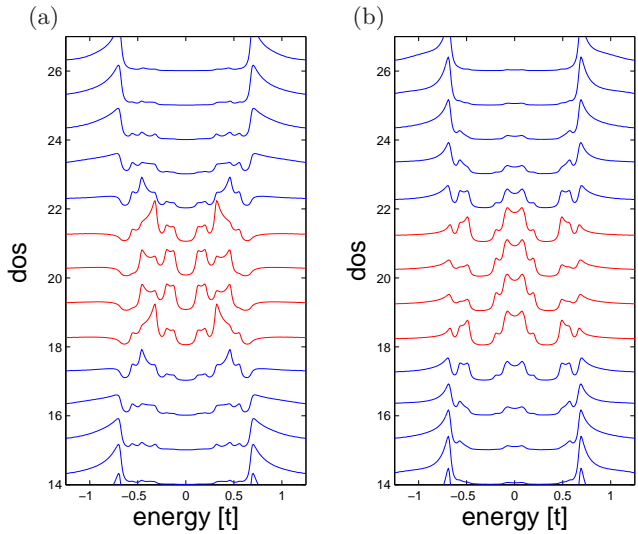}
\caption{(Color online) LDOS for the 100 AF/N/AF (a) and
AF/N/$\pi$AF (b) junction with $d=4$, $\mu=0$, and $U=2.7t$. The
four center LDOS scans are in the N
region.\label{ldos100AFNAFjunction}}
\end{figure}

In Fig. \ref{bandsAFNAF} we plot the eigenbands for the (100)
AF/N/AF junction when $\mu=0$, $U=2.7t$, $d=4$ (a) and $U=2.0t$,
$d=48$ (b). In the limit of a long normal region, we see that
outside the parabola-shaped region centered at $k_y=0.0$, the
bands roughly approach a set of equidistant sine-shaped bands, in
agreement with Eq.(\ref{equi}) at $\theta=0$ for the (100)
orientation. In this limit the low-energy LDOS in the N region
(not shown) displays the expected equally spaced peaks. Note, that
inside the parabola region we observe a qualitatively new
dispersive behavior of the eigenenergies with maximum at $k_y=0$.
We expect these deviations from the simple result of the
quasiclassical treatment to be caused by the assumption $m \ll t$
used to obtain Eq.(\ref{equi}). In addition, the details of the
spectrum near points where $v_{F,x}$ approaches zero can always
differ from the quasiclassical result.

It is clear from Fig. \ref{bandsAFNAF}a that we can have very
strong deviations from the quasiclassical result when $d \sim a$
and $m \sim t$. In particular, within the quasiclassical framework
the dispersion of the bound state energies is entirely associated
with the momentum dependence of the Fermi velocity
$v_{F,x}(\bm{k}_F)$. In contrast, Fig. \ref{bandsAFNAF}a displays
a more complicated momentum dependence of the energy levels with
numerous extrema and additional peaks in the associated LDOS.

\begin{figure}[!tbh]
\includegraphics[width=8.5cm]{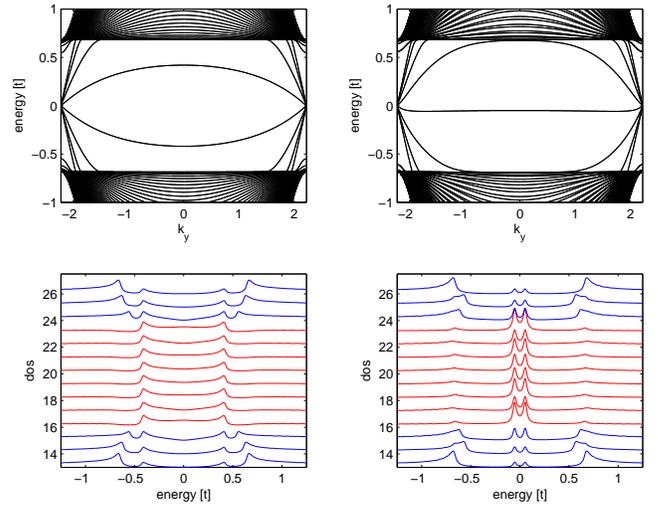}
\caption{(Color online) Eigenbands and the associated LDOS for the
110 AF/N/AF (left column) and 110 AF/N/$\pi$AF (right column)
junction with $d=8$, $\mu=0$, and $U=2.7t$. Note the low-energy
state generated in the spin-$\pi$ junction. In the LDOS plots, the
middle eight scans are in the N region whereas the upper and lower
three shows the LDOS upon moving into the AF.
\label{bandsandldos110AFNAFjunctions}}
\end{figure}

We have also studied a {\sl spin-pi junction}, a AF/N/$\pi$AF
junction with $\theta=\pi$, where the magnetization has gained an
extra $\pi$ phase shift when crossing the normal region N. This is
similar to the $\pi$-phase shift in the stripe phase of the
cuprate superconductors\cite{tranquada,ostripes,andersen05}. For
the AF/N/$\pi$AF junction the subgap bands are shifted compared to
the AF/N/AF configuration. As a consequence, the associated LDOS
will reveal whether the junction exhibits the $\pi$ phase shift or
not. This can be seen in Fig. \ref{ldos100AFNAFjunction} which
shows the LDOS associated with Fig. \ref{bandsAFNAF}a and the
corresponding AF/N/$\pi$AF junction in Fig.
\ref{ldos100AFNAFjunction}b. The surprisingly large number of
peaks seen in Fig. \ref{ldos100AFNAFjunction} agree with the
strongly dispersive bands as can be verified from the stationary
points in Fig. \ref{bandsAFNAF}a. The presence of low-energy
states seen in the AF/N/$\pi$AF junction in Fig.
\ref{ldos100AFNAFjunction}b correlates with predictions of the
quasiclassical approach, which, however, does not apply directly
to the case of very thin normal metal region $d=4$.

Fig. \ref{bandsandldos110AFNAFjunctions} shows the bands and LDOS
for the 110 AF/N/AF (left column) and AF/N/$\pi$AF (right column)
junction, respectively. In accordance with Eqs. (\ref{En}),
(\ref{equi}), the orientational dependence of the spectrum is
associated with the Fermi velocity dispersion. Indeed, in the
limit of large $d$, the bound state spectrum consists again of
equidistant states like in the (100) case. For the 110 geometry we
also find that the main qualitative difference between the AF/N/AF
and AF/N/$\pi$AF junctions is associated with the presence of
almost zero energy states in the latter case.

\section{Conclusions}
We have performed a theoretical study of interfaces and junctions
involving antiferromagnets and superconductors or normal metals.
This was presented both in the framework of quasiclassics and
self-consistent numerical solutions of the relevant Bogoliubov-de
Gennes equations. Where comparison is appropriate, we found full
agreement between the two methods. In particular, we investigated
for formation of bound states near (100) and (110) interfaces
between antiferromagnets and $s$- or $d$-wave superconductors. We
calculated their dispersion, their influence on the proximity
effect, and the associated modifications of the LDOS near the
interface regions. In addition we discussed the crossover between
Q-reflection and conventional specular reflection as a function of
the potential barrier at the interface. In future work we plan to
investigate the role of the bound states on the Josephson current
in SC/AF/SC junctions where the low-energy bound states can be
expected to generate e. g. unusual temperature dependence of the
critical current.

{\it Acknowledgments.} This work was supported by grant
NSF-INT-0340536 (I.V.B., P.J.H., and Yu.S.B.), DOE
DE-FG02-05ER46236 (PJH and Yu.S.B.), and by ONR grant
N00014-04-0060 (P.J.H and B.M.A). I.V.B. and Yu.S.B. also
acknowledge the support by grant RFBR 05-02-15175.


\end{document}